\documentclass[aps,reprint,longbibliography,superscriptaddress,floatfix]{revtex4-2}

\usepackage[T1]{fontenc}
\usepackage[utf8]{inputenc}
\usepackage{amsmath,amssymb,amsfonts,bm,mathtools}
\usepackage{physics}
\usepackage{graphicx}
\usepackage{xcolor}
\usepackage{booktabs}
\usepackage{multirow}
\usepackage{hyperref}
\usepackage{enumitem}
\usepackage{comment}
\usepackage{ragged2e}

\usepackage{orcidlink}

\usepackage{tikz}

\hypersetup{colorlinks=true,linkcolor=blue,citecolor=blue,urlcolor=blue}

\newcommand{\Var}{\operatorname{Var}}
\newcommand{\Cov}{\operatorname{Cov}}

\newcommand{\e}{\mathrm{e}}
\newcommand{\st}{\mathrm{st}}

\graphicspath{{figures/}}
\begin{document}
	
	\title{Finite-size reliability of homothetic quantum Otto engines}
	\author{Gabriella G. Damas~\orcidlink{0000-0003-3376-9281}}
	\affiliation{Department of Physics, Zhejiang Normal University, Jinhua 321004, China}
	\affiliation{Instituto de Física, Universidade Federal de Goiás, 74.001-970, Goiânia
		- GO, Brazil}

    \author{Clebson Cruz~\orcidlink{0000-0003-3318-1111}}
    \affiliation{Centro das Ciências Exatas e das Tecnologias, Universidade Federal 
    do Oeste da Bahia, Rua Bertioga 892, Morada Nobre, 47810-059}
	
	\author{Norton G. de Almeida~\orcidlink{0000-0001-8517-6774}}
	\affiliation{Instituto de Física, Universidade Federal de Goiás, 74.001-970, Goiânia
		- GO, Brazil}
	
	\author{Gao Xianlong~\orcidlink{0000-0001-6914-3163}}
	\email{gaoxl@zjnu.edu.cn}
	\affiliation{Department of Physics, Zhejiang Normal University, Jinhua 321004, China}
	
	\author{G. D. de Moraes Neto~\orcidlink{0000-0003-4273-8380}}
	\email{gdmneto@gmail.com}
	\affiliation{Department of Fundamental Sciences, Hainan Bielefeld University of Applied Sciences, Danzhou, Hainan 578101, China}

\begin{abstract}
Homothetic quantum Otto engines---where all populated energy gaps are rescaled by a common factor---provide a reference model in which the quasistatic stochastic efficiency is trajectory‑independent while work remains fluctuating. For arbitrary finite homothetic spectra we derive the two-point-measurement work distribution and reduce the first two work moments to endpoint energy moments. Specializing to a uniformly spaced ladder gives closed finite-$N$ expressions for the full work distribution, mean work, variance, and signal-to-width reliability. This ladder connects the qubit and oscillator limits, reveals a finite-$N$ reliability crossover, and demonstrates that the high-temperature and infinite-dimensional limits do not commute. The noncommutation reflects a bounded-versus-unbounded spectral distinction: at fixed finite $N$ the Gibbs state has a normalizable infinite-temperature limit, whereas the oscillator retains an ever-expanding thermal tail. The exact formulas are used to compare standard mean-output prescriptions with work reliability, showing that maximum mean output and maximum dimensionless reliability select different operating points. The benchmark is extended to incomplete diagonal reset and to finite-time unitary strokes described by transition matrices, with a finite-ladder protocol and a harmonic sudden-switch oscillator benchmark as controlled examples. Weak deviations from exact homothety are treated perturbatively, showing how level-dependent gap distortions reintroduce quasistatic efficiency fluctuations and modify work reliability. Together, these results separate finite-size, incomplete thermalization, finite-time, and weak spectral-distortion contributions to work unreliability in quantum Otto engines.
\end{abstract}

	\maketitle
\section{Introduction}

Microscopic heat engines are inherently noisy: the work they extract from cycle to cycle fluctuates, and these fluctuations can dominate when only a few energy levels are thermally occupied. Quantum heat engines formulate thermodynamic cycles in terms of discrete spectra, driven Hamiltonians, thermalization strokes, and microscopic energy exchanges~\cite{Scovil1959,Alicki1979,Kosloff1984,Kieu2004,Quan2005,Quan2007,Kosloff2013,Goold2016,Vinjanampathy2016,DeffnerCampbell2019,Cangemi2024}. The quantum Otto cycle is a standard benchmark because its operation separates into two unitary work strokes and two isochoric thermalization strokes~\cite{RezekKosloff2006,KosloffRezek2017}. This structure has been studied for qubits, harmonic oscillators, spins, multilevel systems, interacting working media, and critical systems, and has been demonstrated in trapped ions, nuclear-spin platforms, defect centers, collisional reservoirs, and superconducting circuits~\cite{Abah2012,Rossnagel2016,Peterson2019,Klatzow2019,Ono2020,Bouton2021,Uusnakki2026,XuHe2024,Xu2024AQRSM}. In such microscopic engines, mean work and mean efficiency do not by themselves characterise operation; the fluctuations of work, heat, and efficiency determine how reliably work can be extracted from one cycle to the next.

The spectral class considered here is fixed by \emph{common gap rescaling}. Quan \emph{et al.} showed that a quantum Carnot cycle is thermodynamically reversible only when the quantum adiabatic strokes map Gibbs populations at one bath temperature into Gibbs populations at the other bath temperature; for a multilevel working medium this requires all energy gaps to be rescaled by a common factor~\cite{Quan2007}. We call such a spectrum \emph{homothetic}: every occupied transition gap of the low Hamiltonian is proportional to the corresponding gap of the high Hamiltonian, $E_n^l-E_m^l = \alpha (E_n^h-E_m^h)$ for all $n,m$, with a common scale factor $0<\alpha<1$. In the reversible Carnot connection, this ratio is fixed by the bath temperatures; in an Otto cycle it is a control parameter subject to the engine-operation condition. This class encompasses the two standard Otto boundaries---the two-level engine, whose efficiency is set by a single gap ratio~\cite{Kieu2004,Quan2007}, and the harmonic oscillator engine, whose efficiency is set by a frequency ratio~\cite{RezekKosloff2006,KosloffRezek2017}---and also applies to finite ladders, weakly anharmonic qudits, oscillator cutoffs, and scale-invariant many-body working media~\cite{Jaramillo2016}. Homothety freezes the quasistatic stochastic efficiency to a fixed value, but it does \emph{not} make work deterministic. The work output still fluctuates because the two isochores independently sample thermal energy labels.

At the trajectory level, work, heat, efficiency, and power are stochastic variables. Their relative fluctuations are especially relevant when a small number of levels dominate the cycle statistics. We use the two-point-measurement (TPM) construction for work statistics, the standard operational framework for quantum fluctuation relations, which yields a positive probability distribution for initially energy-diagonal states~\cite{Talkner2007,Esposito2009,Campisi2011,Solinas2015,Campisi2015,Batalhao2014}. Stochastic efficiency and reliability have been studied through efficiency distributions, large-deviation functions, finite-time efficiency statistics, thermodynamic uncertainty relations, and power-efficiency-constancy trade-offs in both classical and quantum settings~\cite{Verley2014NC,Verley2014PRE,Gingrich2014,Polettini2015,BaratoSeifert2015,Gingrich2016,Pietzonka2018,Miller2021}. Quantum Otto studies have further addressed efficiency fluctuations, finite-time irreversible fluctuations, and work/efficiency reliability in multilevel engines~\cite{DenzlerLutz2020,DenzlerLutz2021NJP,Jiao2021,Fei2022,Anka2024,SaryalAgarwalla2021,ShastriVenkatesh2024,McKeever2025}. The TPM treatment used here assumes coherence-erasing isochores that prepare energy-diagonal endpoint states. Coherent engines, dynamic-Bayesian-network formulations, and nonequilibrium reservoirs require additional stochastic variables and can produce work or heat statistics that differ from TPM statistics because projective energy measurements remove coherence~\cite{Micadei2020QFTBeyondTPM,Micadei2021DBNExperimental,RodriguesLutz2024Coherence,damas2026coherence}.

Because homothety eliminates quasistatic efficiency fluctuations, this class provides an ideal null model in which to study work reliability. Once the efficiency is frozen, the remaining TPM work distribution isolates the role of finite Hilbert-space support. The relevant crossover is the finite-$N$ work-reliability crossover from a qubit, through a finite uniform ladder, to the oscillator limit. As we will show, the finite-ladder and oscillator limits behave strikingly differently at high temperatures---they do not commute. This means that even a very large but strictly finite engine can behave qualitatively unlike a true oscillator, with direct consequences for reliability estimates.

The qubit and oscillator boundaries have separate optimization literatures. For two-level Otto machines, maximum-power, ecological, refrigerator, and entropy-production criteria select different operating points~\cite{SinghAbah2020}. For harmonic-oscillator Otto engines and refrigerators, finite-time performance, sudden frequency switching, quantum friction, shortcut-to-adiabaticity protocols, and $\Omega$-function optimization are standard benchmarks~\cite{RezekKosloff2006,Abah2012,DelCampo2014,Alecce2015,AbahLutz2016,Singh2022Unified}. Those works primarily optimize mean thermodynamic performance. A complementary---and less explored---question is how concentrated the work and power outputs remain once an operating point has been selected, and how much stochastic-efficiency spread is generated by finite-time transitions.

While stochastic efficiency statistics, fluctuation bounds, and finite-time effects in scale-invariant engines have been studied before, the exact finite-$N$ work distribution and the reliability crossover connecting the qubit and oscillator boundaries of the same homothetic class have not been obtained. This is the gap we fill. We start with arbitrary finite homothetic spectra and show that the first two work moments reduce to endpoint energy moments without assuming equally spaced levels. Specializing to a uniformly spaced finite ladder gives explicit formulas for the full discrete work distribution, cumulant-generating function, mean, variance, and signal-to-width reliability. These expressions recover the qubit and oscillator mean-work limits while resolving the finite-$N$ work fluctuations between them. They also uncover a low-temperature effective two-level regime, a finite-$N$ high-temperature regime, and the noncommutation of the high-temperature and infinite-dimensional limits. From this noncommutation we extract a useful-dimension criterion: the number of accessible levels required for the finite ladder to behave as an oscillator within a chosen tolerance.

The high-temperature endpoint of a finite ladder is qualitatively different from that of an oscillator. For fixed finite $N$, both endpoint Gibbs states approach the same maximally mixed state on a bounded support. In contrast, the oscillator has no normalizable Gibbs state at $\beta=0$; high temperature instead expands the thermally occupied tail. This order‑of‑limits test reveals when a finite working medium can safely be treated as an oscillator in fluctuation diagnostics.

After establishing the basic benchmark, we evaluate reliability at standard optimized operating points and show that maximum output and maximum reliability generally select different parameters. Complete thermalization is then replaced by a phenomenological diagonal partial-reset map, giving the stationary-cycle work distribution for that channel and isolating diagonal athermality as a separate reliability penalty. Finite-time unitary strokes are included through transition matrices. Once a stroke model supplies those matrices, finite-time drift and jump-spread corrections can be separated from the underlying quasistatic finite-size work fluctuations; a nearest-neighbor protocol serves as an explicit finite-$N$ example. Finally, weak deviations from exact homothety are treated perturbatively, showing how level-dependent gap distortions reintroduce quasistatic efficiency fluctuations. Throughout, the framework cleanly disentangles the contributions from finite size, incomplete thermalization, finite-time driving, and spectral non‑homothety to the work unreliability. The exact formulas we derive are directly applicable to current experimental platforms using qudits, superconducting circuits, or trapped ions, and serve as a diagnostic benchmark for future heat‑engine characterizations.

\section{Homothetic Otto cycle and trajectory-level work}
\label{sec:cycle}

The reference cycle is the quasistatic homothetic Otto cycle~\cite{Quan2005}. The working medium has two externally controlled Hamiltonians: a high-frequency Hamiltonian $H_h$, with spectrum $\{E_n^h\}_{n=0}^{N-1}$, and a low-frequency Hamiltonian $H_l$, with spectrum $\{E_n^l\}_{n=0}^{N-1}$. The cycle consists of a hot isochore, an adiabatic expansion from $H_h$ to $H_l$, a cold isochore, and an adiabatic compression from $H_l$ back to $H_h$. In the quasistatic adiabatic limit, the unitary strokes preserve the energy-level label. Stochasticity enters through the thermal energy labels sampled by the two isochores.

The two-point-measurement population picture is used for energy-diagonal endpoint states~\cite{Talkner2007}. After the hot isochore, the label $n$ is sampled from the Gibbs distribution of $H_h$,
\begin{equation}
p_n^h
=
\frac{e^{-\beta_h E_n^h}}{Z_h^{(N)}},
\qquad
Z_h^{(N)}
=
\sum_{n=0}^{N-1}e^{-\beta_h E_n^h}.
\label{eq:ph_def}
\end{equation}
After the cold isochore, the label $m$ is sampled from the Gibbs distribution of $H_l$,
\begin{equation}
p_m^l
=
\frac{e^{-\beta_l E_m^l}}{Z_l^{(N)}},
\qquad
Z_l^{(N)}
=
\sum_{m=0}^{N-1}e^{-\beta_l E_m^l}.
\label{eq:pl_def}
\end{equation}
Because each isochore fully resets the working medium to a thermal state, and the subsequent unitary stroke does not introduce correlations with the previous isochore, the two endpoint samples $n$ and $m$ are statistically independent. A quasistatic trajectory is therefore completely specified by the pair $(n,m)$: $n$ determines the expansion stroke and $m$ the compression stroke.

The spectral class considered here is defined by the homothetic gap condition
\begin{equation}
E_n^l-E_m^l
=
\alpha\left(E_n^h-E_m^h\right),
\qquad
0<\alpha<1,
\label{eq:homothetic}
\end{equation}
for all allowed labels $n,m$---that is, every gap present in the finite spectrum. Equivalently, the spectra are related by the affine map
\begin{equation}
E_n^l
=
\alpha E_n^h+\chi,
\label{eq:affine}
\end{equation}
where $\chi$ is an overall constant shift that has no physical consequences: it drops out of all normalized Boltzmann weights and of any work or heat difference. One may therefore set $\chi=0$ without loss of generality. This is common gap rescaling for a finite spectrum: all transition gaps are multiplied by the same factor $\alpha$. Appendix~\ref{app:quan} gives the short derivation linking this condition to adiabatic Gibbs-to-Gibbs preservation. In the Otto cycle considered here, $\alpha$ is a free operating parameter; the reversible Carnot-matching value is a special case.

Equations~\eqref{eq:homothetic} and \eqref{eq:affine} do not require uniformly spaced levels. The arbitrary-spectrum formulas derived below apply to any finite discrete spectrum satisfying the homothetic condition, including finite truncations of nonuniform scale-invariant spectra when the level ordering is preserved. The uniform ladder, introduced in Sec.~\ref{sec:ladder}, is the analytically solvable case used for the figures and for the qubit-to-oscillator interpolation. For that model one has $E_n^h=n\epsilon_h$, and the homothetic low-energy spectrum is simply $E_n^l = \alpha n\epsilon_h$ (again up to an irrelevant additive constant).

Positive work denotes work extracted from the engine. Along the expansion stroke, a trajectory starting in level $n$ extracts $E_n^h-E_n^l$. Along the compression stroke, a trajectory starting in the cold level $m$ requires work input $E_m^h-E_m^l$. The net extracted work over the two adiabatic strokes is therefore
\begin{equation}
W(n,m)
=
\left(E_n^h-E_n^l\right)
-
\left(E_m^h-E_m^l\right).
\label{eq:Wdef}
\end{equation}
Using Eq.~\eqref{eq:affine}, this simplifies to
\begin{equation}
W(n,m)
=
(1-\alpha)(E_n^h-E_m^h).
\label{eq:Wtraj}
\end{equation}
The same trajectory absorbs the hot heat
\begin{equation}
Q_h(n,m)=E_n^h-E_m^h,
\label{eq:Qh_traj}
\end{equation}
so that on trajectories where $Q_h\neq0$ the stochastic efficiency is
\begin{equation}
\eta_{\rm st}(n,m)
=
\frac{W(n,m)}{Q_h(n,m)}
=
1-\alpha .
\label{eq:eta_st}
\end{equation}
On trajectories with $Q_h=0$ (i.e.\ $n=m$) the efficiency is undefined; we therefore condition all efficiency statistics on the set of engine-like trajectories with $Q_h>0$ and $W>0$, a convention used throughout the paper. Equation~\eqref{eq:eta_st} displays the characteristic simplification of the homothetic Otto class: in the quasistatic adiabatic limit, the stochastic efficiency is trajectory independent. For scale-invariant Otto engines, the collapse of the TPM stochastic-efficiency distribution to the macroscopic Otto value is already known~\cite{DenzlerLutz2020}. Here it serves as the reference point for studying work fluctuations. Homothety removes quasistatic efficiency fluctuations, but it does not remove work fluctuations, because $W(n,m)$ still depends on two independently sampled endpoint energy labels.

\section{Moment reduction and full TPM distribution}
\label{sec:moments}

The trajectory relation in Eq.~\eqref{eq:Wtraj} fixes the complete-reset quasistatic TPM work distribution:
\begin{equation}
P_N(W)
=
\sum_{n,m=0}^{N-1}p_n^h p_m^l\,
\delta\!\left[W-(1-\alpha)(E_n^h-E_m^h)\right].
\label{eq:general_PW}
\end{equation}
The affine relation $E_m^l=\alpha E_m^h+\chi$ allows the cold Gibbs weights to be written over the hot-spectrum energy variable,
\begin{equation}
p_m^l
=
\frac{e^{-\beta_l E_m^l}}
{\sum_{j=0}^{N-1}e^{-\beta_l E_j^l}}
=
\frac{e^{-\beta_l\alpha E_m^h}}
{\sum_{j=0}^{N-1}e^{-\beta_l\alpha E_j^h}},
\label{eq:pl_hot_variable}
\end{equation}
because the additive shift cancels between numerator and denominator. Both endpoint distributions can therefore be regarded as probability measures on the same ordered set of hot-spectrum energies.

The first two work moments reduce to endpoint energy moments. Define
\begin{equation}
\mu_h=\langle E^h\rangle_h,
\qquad
\mu_l^{(h)}=\langle E^h\rangle_l,
\label{eq:mu_definitions}
\end{equation}
where $\langle\cdot\rangle_h$ denotes averaging with $p^h$, while $\langle\cdot\rangle_l$ denotes averaging with $p^l$. The superscript $(h)$ indicates that, because of the affine map, the cold-side average is taken over the hot-spectrum energy variable. The mean work is therefore
\begin{equation}
\langle W\rangle_N
=
(1-\alpha)(\mu_h-\mu_l^{(h)}),
\label{eq:mean_general}
\end{equation}
and the variance, using the independence of the two endpoint samples, is
\begin{equation}
\sigma_{W,N}^2
=
(1-\alpha)^2
\left[
\operatorname{Var}_h(E^h)+\operatorname{Var}_l(E^h)
\right].
\label{eq:var_general}
\end{equation}
These relations follow directly from Eq.~\eqref{eq:general_PW}. Appendix~\ref{app:moment_details} gives the cumulant-generating function and the explicit moment reduction; the generating function also encodes higher-order cumulants that will be used in later sections.

We quantify reliability by the signal-to-width ratio (the inverse of the coefficient of variation). For any stochastic output $X$,
\begin{equation}
\mathcal R_X=\frac{\langle X\rangle}{\sigma_X}.
\label{eq:generic_reliability}
\end{equation}
For the complete-thermalization, quasistatic, exactly homothetic $N$-level benchmark, the work reliability becomes
\begin{equation}
\mathcal R_N
=
\frac{\langle W\rangle_N}{\sigma_{W,N}}
=
\frac{\mu_h-\mu_l^{(h)}}
{\sqrt{\operatorname{Var}_h(E^h)+\operatorname{Var}_l(E^h)}}.
\label{eq:reliability_general}
\end{equation}
The common work scale $(1-\alpha)$ cancels. This cancellation is special to the complete-thermalization homothetic benchmark. In the extensions below, incomplete thermalization, finite-time transitions, and nonhomothetic spectral distortions modify the trajectory distribution; the same reliability definition is then used with explicit labels such as $\mathcal R_{N,\lambda}$, $\mathcal R_W$, or $\mathcal R_P$.

Because the common work scale factors out of both the mean and the standard deviation, $\mathcal R_N$ is independent of the absolute output magnitude. A large reliability near the zero-output limit $\alpha\to1$ does \textbf{not} by itself signal a useful engine; it simply means that the vanishingly small work has a narrow relative spread. A physically meaningful operating point must therefore be assessed together with a non-zero mean work, power, or another output scale.

\section{Uniform ladder: finite-$N$ solution}
\label{sec:ladder}

\begin{figure*}[!htbp]
\centering
\includegraphics[width=0.75\linewidth]{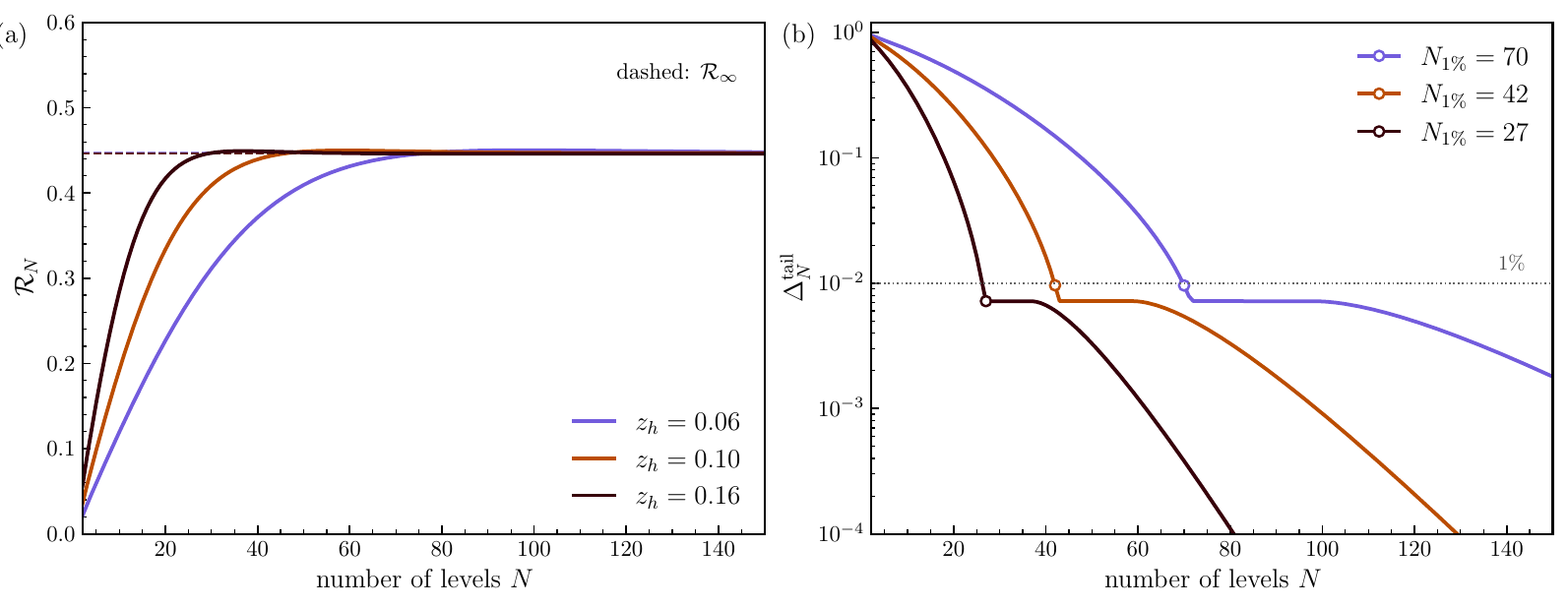}
\caption{
Finite-$N$ crossover of the work reliability $\mathcal R_N$ for a uniformly spaced homothetic ladder.
(a) Exact finite-$N$ reliability as a function of the ladder dimension $N$ for three values of $z_h$; dashed horizontal lines denote the corresponding oscillator limits $\mathcal R_\infty$.
(b) Tail distance $\Delta_N^{\rm tail}=\max_{M\ge N}|\mathcal R_M-\mathcal R_\infty|/\mathcal R_\infty$.
The dotted line marks the $1\%$ tolerance, and the open markers identify the smallest dimension $N_{1\%}$ after which the finite ladder remains within this tolerance.
The ratio of scaled gaps is fixed at $r=z_l/z_h=2.0$.
}
\label{fig:snr_crossover}
\end{figure*}

To explicitly evaluate the finite-size work statistics we now focus on a uniformly spaced finite ladder,
\begin{equation}
E_n^h=n\epsilon,
\qquad
n=0,1,\ldots,N-1,
\qquad
\epsilon=\hbar\omega_h .
\label{eq:uniform_ladder}
\end{equation}
The homothetic low spectrum has spacing $\alpha\epsilon$, up to an additive shift. The work depends only on the index difference $k=n-m$,
\begin{equation}
W_k=(1-\alpha)\epsilon k,
\qquad
k=-(N-1),\ldots,N-1 .
\label{eq:Wk_uniform}
\end{equation}

Because the endpoint thermal labels are sampled independently after the isochores, the work-index distribution reduces to the finite convolution
\begin{equation}
P_N(k)=
\sum_{m=\max(0,-k)}^{\min(N-1,N-1-k)}
p_{m+k}^h p_m^l .
\label{eq:PNk}
\end{equation}

For compactness, we introduce the dimensionless thermal parameters
\begin{equation}
z_h=\beta_h\epsilon,
\qquad
z_l=\beta_l\alpha\epsilon .
\label{eq:zh_zl_def}
\end{equation}
We keep $k_B$ explicit when discussing experimental units and set $k_B=1$ in dimensionless formulas. Thus $z_h$ and $z_l$ are dimensionless inverse temperatures. Using $hf/k_B\simeq 47.99\,{\rm mK}(f/1\,{\rm GHz})$, a $5\,{\rm GHz}$ transition corresponds to $hf/k_B\simeq 240\,{\rm mK}$. Hence $z=1$ corresponds to $T\simeq240\,{\rm mK}$, while $z=0.1$ corresponds to approximately $2.4\,{\rm K}$.

For a finite ladder, all thermodynamic properties are encoded in the canonical partition function
\begin{equation}
Z_N(z)=\sum_{n=0}^{N-1}e^{-zn}
=
\frac{1-e^{-Nz}}{1-e^{-z}} .
\label{eq:ZN}
\end{equation}
The statistical moments of the energy distribution are naturally generated by derivatives of the dimensionless Massieu potential $\Phi=\ln Z_N(z) = -\beta \mathcal{F}$, where $\mathcal{F}$ is the Helmholtz free energy~\cite{Hoyuelos2025}. Preserving the exact finite-$N$ dependence of this potential is essential because its non‑extensive contributions strongly influence thermal fluctuations in a microscopic system~\cite{Hoyuelos2025}. The first moment gives the mean excitation number,
\begin{equation}
\nu_N(z)
=
-\partial_z\ln Z_N(z)
=
\frac{1}{e^z-1}-\frac{N}{e^{Nz}-1},
\label{eq:nuN}
\end{equation}
while the second moment defines the index variance,
\begin{equation}
v_N(z)
=
\partial_z^2\ln Z_N(z)
=
\frac{e^z}{(e^z-1)^2}
-
\frac{N^2e^{Nz}}{(e^{Nz}-1)^2}.
\label{eq:vN}
\end{equation}
Physically, $v_N(z)$ measures the amplitude of thermal fluctuations in the finite-dimensional working medium. Through the fluctuation–dissipation theorem, these equilibrium fluctuations govern the linear response of the system~\cite{Scandi2019,Giacomo2024}. Thus $v_N(z)$ acts as a thermal susceptibility: it quantifies the linear response of the mean excitation number to changes in the dimensionless inverse temperature, and it is directly related to the isochoric heat capacity $\mathcal{C}$ via
\begin{equation}
    v_N(z) = \frac{\mathcal{C}}{k_B z^2} .
    \label{eq:heat_capacity}
\end{equation}

Substituting the equilibrium moments into the general expressions for the work moments connects the work output to the fundamental response functions of the working medium. The mean extracted work becomes
\begin{equation}
\langle W\rangle_N
=
(1-\alpha)\epsilon[\nu_N(z_h)-\nu_N(z_l)],
\label{eq:mean_uniform}
\end{equation}
which shows that the average is simply the quantum of work $W_1=(1-\alpha)\epsilon$ weighted by the net change in the average excitation number between the hot and cold isochores.

Because complete thermalization erases correlations between the beginning and the end of the cycle, the work variance is strictly additive. It is the sum of the independent thermal fluctuations at the endpoints,
\begin{equation}
\sigma_{W,N}^2
=
(1-\alpha)^2\epsilon^2[v_N(z_h)+v_N(z_l)].
\label{eq:var_uniform}
\end{equation}
Consequently, from Eq.~\eqref{eq:heat_capacity}, the work fluctuations are bounded by the thermal capacities of the finite-size working medium.

The work reliability is therefore
\begin{equation}
\mathcal R_N
=
\frac{\nu_N(z_h)-\nu_N(z_l)}
{\sqrt{v_N(z_h)+v_N(z_l)}} .
\label{eq:snr_uniform}
\end{equation}
Positive average work requires $z_h<z_l$, equivalently $T_h>(\omega_h/\omega_c)T_l$ when $\alpha=\omega_c/\omega_h$.

Equation~\eqref{eq:snr_uniform} shows that reliability results from a competition between the macroscopic population displacement $\nu_N(z_h)-\nu_N(z_l)$ and the microscopic thermal susceptibility $\sqrt{v_N(z_h)+v_N(z_l)}$. Highly reliable operation demands maximizing the population transfer while keeping the isochoric heat capacities—and thus the thermal fluctuations—as small as possible.

The Otto efficiency is fixed by the scale factor $1-\alpha$, whereas $\mathcal R_N$ is fixed by the separation of two truncated Gibbs distributions relative to their combined width. Figure~\ref{fig:snr_crossover} shows the resulting finite-$N$ crossover. The qubit boundary is strongly fluctuation limited; increasing $N$ opens the upper thermal tail and moves the reliability toward the oscillator value. The tail-distance diagnostic in panel (b) turns this approach into a useful-dimension criterion.

The finite work distribution also contains non-Gaussian information beyond its width. Just as the exact finite-$N$ Massieu potential governs the equilibrium thermal noise through its non‑extensive contributions~\cite{Hoyuelos2025}, the complete work statistics are captured by the cumulant-generating function of the work index, $K_N(t)=\ln\langle e^{tk}\rangle$. Its first two cumulants reproduce the mean and variance of Eq.~\eqref{eq:snr_uniform}, and its third cumulant defines the normalized skewness
\begin{equation}
\gamma_{1,N}
=
\frac{\kappa_3^{(k)}}{[v_N(z_h)+v_N(z_l)]^{3/2}} .
\label{eq:gamma1_uniform}
\end{equation}
Appendix~\ref{app:ladder_details} gives the explicit forms of $K_N(t)$ and $\kappa_3^{(k)}$. While the reliability $\mathcal R_N$ measures the work signal relative to the width, $\gamma_{1,N}$ quantifies the asymmetry of the TPM work distribution. Physically, this skewness reveals how the finite-size constraints bias the thermal fluctuations toward values larger or smaller than the mean.

The connection between thermal noise and finite system size becomes particularly transparent in the high-temperature regime. As $z\to0$ the working medium approaches a maximally mixed state on its finite support, since the thermal energy greatly exceeds the total energy stored in the ladder. The thermodynamic quantities are defined by continuity:
\[
Z_N(0)=N,
\qquad
\nu_N(0)=\frac{N-1}{2},
\qquad
v_N(0)=\frac{N^2-1}{12}.
\]
This is the saturation regime where the engine loses its ability to resolve the discrete energy structure. As $N$ increases, the work-fluctuation variance grows quadratically, $v_N\sim\mathcal{O}(N^2)$, indicating that at high temperatures the reliability $\mathcal R_N$ [Eq.~\eqref{eq:snr_uniform}] is fundamentally suppressed by the increased thermal susceptibility of larger Hilbert spaces.

\begin{figure*}[!t]
\centering
\includegraphics[width=1\linewidth]{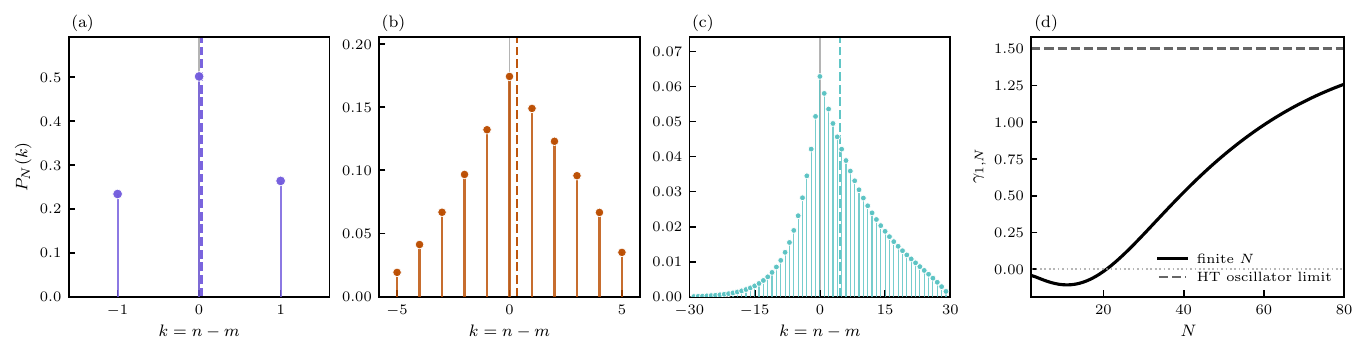}
\caption{
Non-Gaussian finite-size structure of the TPM work distribution for a uniform homothetic ladder with $z_h=0.08$ and $z_l=0.20$.
Panels (a)--(c) show the distribution of the dimensionless work index $k=W/[(1-\alpha)\epsilon]=n-m$ for $N=2$, $N=6$, and $N=30$, respectively. Dashed vertical lines mark $\langle k\rangle_N$.
Panel (d) shows the normalized skewness $\gamma_{1,N}$ from Eq.~\eqref{eq:gamma1_uniform}; the dashed horizontal line gives the oscillator high-temperature limit in Eq.~\eqref{eq:gamma1_HT_osc}.
Finite spectral support changes both the width and the non-Gaussian tail structure of the work distribution.
}
\label{fig:work_distribution}
\end{figure*}

For fixed finite $N$, the strict high-temperature limit makes the two endpoint distributions uniform on the same bounded support. In this regime finite-size truncation effects dominate and rigidly constrain the state space~\cite{Hoyuelos2025}, perfectly symmetrizing the thermal noise so that $\gamma_{1,N}\to0$. If, instead, the oscillator limit is taken first, the geometric thermal tail remains, evading the finite-size cutoff. For $r=z_l/z_h>1$, the oscillator high-temperature skewness approaches
\begin{equation}
\gamma_{1,\infty}^{\rm HT}
=
\frac{2(1-r^{-3})}{(1+r^{-2})^{3/2}} .
\label{eq:gamma1_HT_osc}
\end{equation}
Figure~\ref{fig:work_distribution} shows the same finite-size crossover at the level of the full work-index distribution. Panels (a)--(c) compare $N=2$, $N=6$, and $N=30$, showing how the upper thermal tail develops as the ladder dimension increases. Panel (d) shows the corresponding skewness: finite support drives the distribution away from the oscillator high-temperature value at intermediate dimensions, before the oscillator tail is recovered.

\section{Boundary cases and noncommuting limits}
\label{sec:limits}

Equation~\eqref{eq:snr_uniform} contains several limiting cases that highlight the competition between the state-space dimension and thermal fluctuations. For the most restricted case, $N=2$ (a qubit), the excited-state probabilities are
\begin{equation}
p_h=\frac{1}{1+e^{z_h}},
\qquad
p_l=\frac{1}{1+e^{z_l}}.
\label{eq:qubit_probs}
\end{equation}
The macroscopic work and reliability then read
\begin{align}
\langle W\rangle_2
&=(1-\alpha)\epsilon(p_h-p_l),
\nonumber\\
\mathcal R_2
&=
\frac{p_h-p_l}
{\sqrt{p_h(1-p_h)+p_l(1-p_l)}}.
\label{eq:qubit_results}
\end{align}
Note that the factors $p(1-p)$ are precisely the variances of a Bernoulli distribution, corresponding to the two-level thermal susceptibility $v_2(z)$.

At high temperature the available thermal energy exceeds the single energy gap, forcing the system into a maximally mixed state where state-space saturation rigidly restricts any further energy absorption. Expanding the reliability gives $\mathcal R_2\simeq (z_l-z_h)/(2\sqrt{2})$, so the two-level reliability vanishes linearly. The mean work reduces to the standard qubit Otto expression
\begin{equation}
\langle W\rangle_2
=
\hbar(\omega_h-\omega_c)
\left[
\frac{1}{1+e^{\beta_h\hbar\omega_h}}
-
\frac{1}{1+e^{\beta_l\hbar\omega_c}}
\right],
\label{eq:qubit_standard_mean}
\end{equation}
which is the familiar reference point for two-level Otto optimization and experimental benchmarks~\cite{Kieu2004,SinghAbah2020,Peterson2019}.

The low-temperature regime is also effectively two-level. Because the thermal energy is too small to populate the higher levels, the finite-size truncation contributions to the Massieu potential $\ln Z_N(z)$ remain inactive~\cite{Hoyuelos2025}. For $z_h,z_l\gg1$ and any $N\ge2$, the reliability asymptotically simplifies to
\begin{equation}
\mathcal R_N^{\rm low\,T}
\simeq
\frac{e^{-z_h}-e^{-z_l}}{\sqrt{e^{-z_h}+e^{-z_l}}}.
\label{eq:lowT_snr}
\end{equation}
Higher levels affect the leading work statistics only after they acquire appreciable thermal weight; the finite-$N$ reliability crossover is therefore an intermediate- and high-temperature effect.

The opposite regime is the near‑uniform finite‑$N$ limit. For fixed $N$ and extremely high temperatures such that $N z_h, N z_l \ll 1$, the thermal energy effectively dominates the entire finite spectrum. Expanding Eq.~\eqref{eq:snr_uniform} in this near‑uniform limit yields
\begin{equation}
\mathcal R_N
\simeq
(z_l-z_h)\sqrt{\frac{N^2-1}{24}}.
\label{eq:near_uniform}
\end{equation}
This growth is limited by the thermal occupation range: Eq.~\eqref{eq:near_uniform} assumes the ladder remains nearly uniformly populated, which requires $N z_h\ll1$ and $N z_l\ll1$. Once $N$ exceeds the thermally occupied range, finite-size truncation effects become negligible; added levels acquire negligible probability, and the finite-dimensional working medium smoothly saturates toward the oscillator boundary.

\begin{figure}[!htbp]
\centering
\includegraphics[width=1\linewidth]{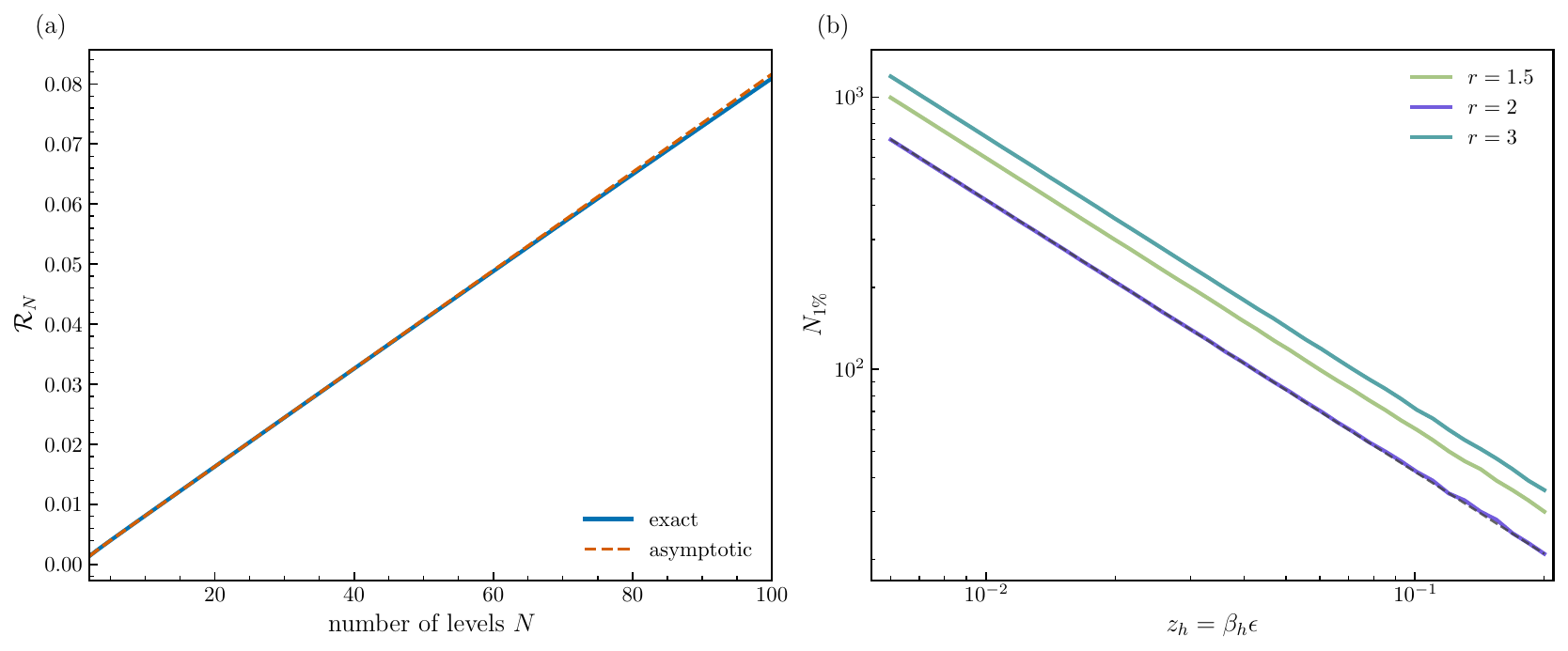}
\caption{
Finite-$N$ scaling and saturation toward oscillator reliability.
(a) Exact work reliability $\mathcal R_N$ compared with the near-uniform expansion $\mathcal R_N^{\rm asy}=(z_l-z_h)\sqrt{(N^2-1)/24}$.
This panel uses $z_h=0.004$, $z_l=0.008$, equivalently $r=z_l/z_h=2.0$.
The agreement confirms the approximately linear finite-$N$ growth in the near-uniform regime.
(b) Minimum dimension $N_{1\%}$ required for the finite ladder to remain within $1\%$ of the oscillator reliability, plotted versus $z_h=\beta_h\epsilon$ at fixed ratios $r=z_l/z_h$. The scan uses $z_h\in[0.006,0.20]$ on a logarithmic grid, and $z_l=r z_h$.
}
\label{fig:asymptotic_saturation}
\end{figure}

Figure~\ref{fig:asymptotic_saturation} illustrates this two‑stage behavior. Panel (a) compares the exact reliability with the near‑uniform expansion. Panel (b) converts the saturation into a useful‑dimension criterion: $N_{1\%}$ is the smallest dimension after which the reliability remains within $1\%$ of $\mathcal R_\infty$. The required dimension grows as $z_h$ decreases because the oscillator thermal tail broadens at high temperature. The relevant scale is therefore not $N$ alone, but the product of the accessible dimension and the thermal scale. Appendix~\ref{app:dimension_design} gives design maps for both work reliability and mean work and shows the diminishing returns once the thermally active part of the spectrum has been resolved.

We now examine the continuous thermodynamic limit. Taking $N\to\infty$ at fixed dimensionless temperature $z>0$ completely removes the finite-size constraints. The working medium recovers the Massieu potential of a standard harmonic oscillator, giving the well‑known macroscopic moments
\begin{align}
Z_\infty(z)&=\frac{1}{1-e^{-z}},
\nonumber\\
\nu_\infty(z)&=\frac{1}{e^z-1},
\nonumber\\
v_\infty(z)&=\frac{e^z}{(e^z-1)^2}.
\label{eq:osc_moments}
\end{align}
The oscillator reliability is
\begin{equation}
\mathcal R_\infty
=
\frac{(e^{z_h}-1)^{-1}-(e^{z_l}-1)^{-1}}
{\sqrt{e^{z_h}(e^{z_h}-1)^{-2}+e^{z_l}(e^{z_l}-1)^{-2}}}.
\label{eq:snr_inf}
\end{equation}

\begin{figure*}[!htbp]
\centering
\includegraphics[width=0.8\linewidth]{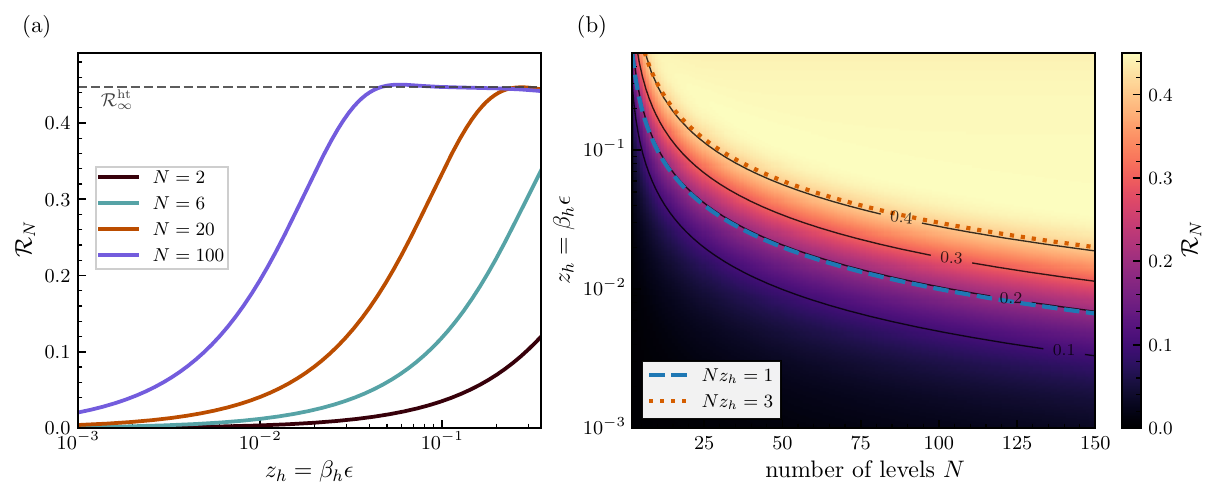}
\caption{
Noncommuting high-temperature and infinite-dimensional limits of work reliability.
(a) $\mathcal R_N$ versus $z_h=\beta_h\epsilon$ at fixed ratio $r=z_l/z_h=2$.
For each finite $N$, $\mathcal R_N\to0$ as $z_h\to0$, whereas the oscillator-first limit gives the plateau $\mathcal R_\infty^{\rm ht}=(r-1)/\sqrt{1+r^2}$.
(b) Dimension-temperature landscape of $\mathcal R_N$ at $r=2$.
The crossover to oscillator-like reliability is governed mainly by $N z_h$; guide curves $N z_h=1$ and $N z_h=3$ indicate the onset of cutoff-insensitive behavior.
}
\label{fig:limits_landscape}
\end{figure*}

The physical divergence between a finite and an infinite state space becomes most apparent in the high‑temperature limit. Expanding the oscillator reliability at high temperatures with a fixed ratio $r=z_l/z_h>1$ gives a non‑zero asymptotic plateau,
\begin{equation}
\mathcal R_\infty
\to
\mathcal R_\infty^{\rm ht}
=
\frac{r-1}{\sqrt{1+r^2}}.
\label{eq:osc_plateau}
\end{equation}
Every fixed finite-$N$ ladder, by contrast, satisfies $\mathcal R_N\to0$ as $z_h\to0$. Hence, at fixed $r=z_l/z_h>1$,
\begin{equation}
\lim_{z_h\to0}\lim_{N\to\infty}\mathcal R_N
= \frac{r-1}{\sqrt{1+r^2}},
\qquad
\lim_{N\to\infty}\lim_{z_h\to0}\mathcal R_N
=
0 .
\label{eq:noncommuting_limits_explicit}
\end{equation}
The high‑temperature and infinite‑dimensional limits therefore do not commute. The origin of this noncommutation is the fundamental difference between bounded and unbounded spectral support, which dictates how the thermal capacity of the system responds to extreme energy fluctuations. At fixed finite $N$, the limit $z_h,z_l\to0$ sends both endpoint Gibbs distributions to the same uniform distribution on $\{0,\ldots,N-1\}$. In this maximally mixed state, finite‑size truncation effects rigidly constrain the state space, and the mean displacement $\nu_N(z_h)-\nu_N(z_l)$ vanishes faster than the finite width can sustain a nonzero signal‑to‑width ratio: the macroscopic work signal collapses while the thermal noise is maximized. If the oscillator limit is taken first, the Gibbs distribution remains normalizable for every $z>0$ but has no normalizable $z=0$ endpoint. The high‑temperature limit then probes an expanding geometric tail, for which both the mean displacement and the width scale as $1/z_h$, leaving the finite plateau of Eq.~\eqref{eq:osc_plateau}. Thus the oscillator high‑temperature limit is not the uniform high‑temperature limit of a finite ladder; it is a joint large‑support/high‑temperature limit. Recent results have explored this same spectral dichotomy in the precision limits of quantum thermometry~\cite{Aiache2026}, where finite‑spectrum probes and unbounded continuous oscillators fall into distinct high‑temperature universality classes. Figure~\ref{fig:limits_landscape} displays the noncommutation through finite‑$N$ curves and a dimension–temperature landscape organized mainly by $N z_h$.

The variable $N z_h$ has a direct cutoff interpretation: it captures the competition between the available thermal energy and the thermodynamic ceiling of the working medium. For the uniform ladder, the upper energy scale is $E_{\max}\simeq N\epsilon$, so $N z_h\simeq\beta_h E_{\max}$. The regime $N z_h\ll1$ is the finite‑support high‑temperature regime: the thermal energy exceeds the available spectral width and the ladder is nearly uniformly populated. The regime $N z_h\gg1$ is cutoff insensitive: the thermally populated tail is resolved before the upper boundary is reached, finite‑size truncation has no observable thermodynamic consequences, and the finite ladder approaches the oscillator result. The useful‑dimension criterion in Fig.~\ref{fig:asymptotic_saturation} is therefore a practical form of the condition $\beta_h E_{\max}\gg1$, which ensures that the working medium operates safely below its saturation threshold.

\subsection{Heat-current and refrigerator interpretation}
\label{subsec:heat_current_limits}

The same order‑of‑limits structure appears in the heat exchanged with the isochores. In the quasistatic homothetic cycle, the hot heat and work are not independent trajectory variables. From Eqs.~\eqref{eq:Wtraj} and~\eqref{eq:Qh_traj},
\begin{align}
W(n,m)&=(1-\alpha)Q_h(n,m), \nonumber\\
Q_h(n,m)&=E_n^h-E_m^h .
\label{eq:W_Qh_relation}
\end{align}
For the uniform ladder this gives
\begin{equation}
Q_h(k)=\epsilon k,
\qquad
W(k)=(1-\alpha)\epsilon k,
\qquad
k=n-m .
\label{eq:Qh_ladder}
\end{equation}
Consequently,
\begin{align}
\langle Q_h\rangle_N
&=
\epsilon[\nu_N(z_h)-\nu_N(z_l)],
\label{eq:Qh_mean_N}
\\
\sigma_{Q_h,N}^2
&=
\epsilon^2[v_N(z_h)+v_N(z_l)] .
\label{eq:Qh_var_N}
\end{align}
If the cycle duration is fixed at $\tau_{\rm cyc}$, the cycle‑averaged hot heat current and its per‑cycle noise scale are
\begin{equation}
J_h^{(N)}=\frac{\langle Q_h\rangle_N}{\tau_{\rm cyc}},
\qquad
S_h^{(N)}=\frac{\sigma_{Q_h,N}^2}{\tau_{\rm cyc}} .
\label{eq:Jh_Sh_def}
\end{equation}
The hot‑heat reliability is therefore identical to the work reliability,
\begin{equation}
\mathcal R_{Q_h,N}
=
\frac{\langle Q_h\rangle_N}{\sigma_{Q_h,N}}
=
\mathcal R_N .
\label{eq:Qh_reliability}
\end{equation}
The raw current shows an even sharper version of the same noncommutation. For every fixed finite $N$, the high‑temperature limit with fixed ratio $r=z_l/z_h>1$ makes the two endpoint distributions uniform and therefore
\begin{equation}
\lim_{z_h\to0}J_h^{(N)}=0,
\qquad
N<\infty .
\label{eq:Jh_finite_HT}
\end{equation}
Taking $N\to\infty$ first gives the oscillator expression
\begin{equation}
\langle Q_h\rangle_\infty
=
\epsilon\left[
\frac{1}{e^{z_h}-1}
-
\frac{1}{e^{z_l}-1}
\right],
\label{eq:Qh_oscillator}
\end{equation}
so that, at fixed $r=z_l/z_h>1$,
\begin{equation}
\langle Q_h\rangle_\infty
\simeq
\frac{\epsilon}{z_h}\left(1-\frac{1}{r}\right),
\qquad
z_h\to0 .
\label{eq:Qh_oscillator_HT}
\end{equation}
Thus
\begin{equation}
\lim_{z_h\to0}\lim_{N\to\infty}J_h^{(N)}
=
+\infty,
\qquad
\lim_{N\to\infty}\lim_{z_h\to0}J_h^{(N)}
=
0 .
\label{eq:Jh_noncommuting}
\end{equation}
The divergence of the oscillator‑first current is a consequence of the unbounded thermal occupation. By contrast, the normalized heat signal remains finite in the oscillator‑first limit because the heat‑current mean and width scale with the same thermal occupation scale.

The cold heat obeys the same algebraic structure. With the sign convention used above, the heat delivered to the cold bath during engine operation is
\begin{equation}
Q_{c,{\rm out}}(n,m)=\alpha\epsilon(n-m),
\label{eq:Qc_out_engine}
\end{equation}
while the heat absorbed from the cold bath in refrigerator operation is
\begin{equation}
Q_c^{\rm abs}(n,m)=\alpha\epsilon(m-n).
\label{eq:Qc_abs_fridge}
\end{equation}
In the quasistatic homothetic refrigerator regime, where $z_l<z_h$, the mean absorbed cold heat and its reliability are
\begin{align}
\langle Q_c^{\rm abs}\rangle_N
&=
\alpha\epsilon[\nu_N(z_l)-\nu_N(z_h)],
\label{eq:Qc_abs_mean}
\\
\mathcal R_{c,N}^{\rm ref}
&=
\frac{\nu_N(z_l)-\nu_N(z_h)}
{\sqrt{v_N(z_h)+v_N(z_l)}} .
\label{eq:fridge_reliability}
\end{align}
The quasistatic coefficient of performance remains geometric,
\begin{equation}
{\rm COP}_{\rm Otto}=\frac{\alpha}{1-\alpha},
\label{eq:otto_COP_geometric}
\end{equation}
whereas the cooling signal and its reliability remain controlled by the finite spectral support. Just as a finite upper bound restricts heat absorption, it also imposes a strict limitation on the maximum amount of entropy that the working medium can extract from the cold reservoir. Hence the same finite‑$N$/oscillator noncommutation that appears in work extraction also appears in cooling reliability.

\subsection{Carnot-matching boundary}

The reversible Carnot‑matching line provides another boundary. In the quantum Carnot construction of Quan \textit{et al.}, common gap rescaling is fixed by the bath temperatures so that the adiabatic strokes map a Gibbs state at $T_h$ into a Gibbs state at $T_l$~\cite{Quan2007}. In the homothetic Otto cycle considered here, the same spectral structure appears, but $\alpha$ remains an operating parameter. With
\begin{equation}
\tau=\frac{T_l}{T_h},
\qquad
\frac{z_l}{z_h}=\frac{\alpha}{\tau},
\label{eq:app_carnot_tau}
\end{equation}
the positive‑work condition $z_l>z_h$ is equivalent to $\alpha>\tau$. The Carnot‑efficiency boundary is approached as $\alpha\to\tau^+$. At the boundary, $z_l=z_h$, the two endpoint Gibbs distributions coincide, the finite‑size Massieu potentials perfectly overlap, and the macroscopic population displacement vanishes while the microscopic thermal susceptibility remains strictly positive. Consequently,
\begin{align}
\langle W\rangle_N
&=
(1-\alpha)\epsilon[\nu_N(z_h)-\nu_N(z_l)]
=
0,
\label{eq:app_carnot_zero_work}
\\
\mathcal R_N &= 0 .
\label{eq:app_carnot_zero_reliability}
\end{align}
The homothetic Otto engine reaches the Carnot‑efficiency boundary only as a zero‑output, zero‑reliability limit. Close to this boundary, with $r=z_l/z_h=\alpha/\tau=1+\delta$,
\begin{equation}
\mathcal R_N
\simeq
z_h\sqrt{\frac{v_N(z_h)}{2}}\,\delta .
\label{eq:app_carnot_reliability_scaling}
\end{equation}
The reliability collapses linearly with the distance from the Carnot‑matching boundary. In the oscillator high‑temperature limit, $\mathcal R_\infty\simeq\delta/\sqrt2$, whereas a fixed finite ladder gives
\begin{equation}
\mathcal R_N
\simeq
z_h\sqrt{\frac{N^2-1}{24}}\,\delta .
\label{eq:app_carnot_reliability_finite_ladder}
\end{equation}
This is the finite‑ladder form of the power–efficiency–constancy trade‑off: approaching reversible efficiency suppresses the work signal, and to overcome the finite‑size truncation fluctuations, finite ladders require a thermally large accessible Hilbert space to retain reliability~\cite{Shiraishi2016,Pietzonka2018,Verley2014PRE}.

The oscillator mean work follows from Eq.~\eqref{eq:mean_uniform} by taking $N\to\infty$:
\begin{equation}
\langle W\rangle_\infty
=
\hbar(\omega_h-\omega_c)
\left[
\frac{1}{e^{\beta_h\hbar\omega_h}-1}
-
\frac{1}{e^{\beta_l\hbar\omega_c}-1}
\right],
\label{eq:oscillator_mean_work}
\end{equation}
which is the standard harmonic Otto result~\cite{RezekKosloff2006,KosloffRezek2017}. The finite‑$N$ formulas contain the usual qubit and oscillator mean outputs as boundary cases and give the corresponding interpolation of work fluctuations.

\section{Mean-output optimization versus work reliability}
\label{sec:optimized_reliability}

Optimizing mean performance fixes an operating point but says nothing about the width of the work distribution. We therefore evaluate the exact finite-ladder work reliability at operating points used in the qubit and harmonic-Otto optimization literature~\cite{SinghAbah2020,Singh2022Unified}. We fix the compression ratio by a standard mean-output prescription and then evaluate the resulting work reliability from the finite-$N$ TPM distribution.

Introduce the bath-temperature ratio
\begin{equation}
\tau=\frac{T_l}{T_h}=\frac{\beta_h}{\beta_l},
\qquad
\eta_C=1-\tau .
\end{equation}
Throughout the optimization, the hot gap $\epsilon=\hbar\omega_h$ is held fixed. Hence $z_h=\beta_h\epsilon$ is fixed, while the cold scaled gap varies with the compression ratio,
\begin{equation}
z_l(\alpha)=\beta_l\alpha\epsilon=\frac{\alpha z_h}{\tau}.
\label{eq:zl_alpha}
\end{equation}
Varying $\alpha$ therefore changes both the Otto efficiency $\eta=1-\alpha$ and the cold endpoint distribution entering the work statistics.

Engine operation requires $\tau<\alpha<1$. For a fixed cycle time, maximizing mean power is equivalent to maximizing mean work. Define
\begin{equation}
\bar w_N(\alpha)
=
(1-\alpha)
\left[
\nu_N(z_h)-\nu_N\!\left(\frac{\alpha z_h}{\tau}\right)
\right],
\label{eq:wbar_alpha}
\end{equation}
with $\langle W\rangle_N=\epsilon\,\bar w_N(\alpha)$. The finite-$N$ maximum-output compression ratio is
\begin{equation}
\alpha_{\rm MP}^{(N)}
=
\arg\max_{\tau<\alpha<1}\bar w_N(\alpha).
\label{eq:alpha_mp_N}
\end{equation}
At any selected value of $\alpha$, the corresponding work reliability is
\begin{equation}
\mathcal R_N(\alpha)
=
\frac{\nu_N(z_h)-\nu_N(\alpha z_h/\tau)}
{\sqrt{v_N(z_h)+v_N(\alpha z_h/\tau)}}.
\label{eq:RN_alpha}
\end{equation}
Equation~\eqref{eq:alpha_mp_N} selects an operating point from a mean-output objective; Eq.~\eqref{eq:RN_alpha} evaluates the relative width of the work distribution at that point. Note that we do not treat $\mathcal R_N$ as an independent optimization target. In the homothetic quasistatic benchmark, the factor $(1-\alpha)$ cancels from $\mathcal R_N$, so a large signal-to-width ratio can occur near low-output regions. Reliability is therefore interpreted together with nonzero mean work or power.

For general $N$, Eq.~\eqref{eq:alpha_mp_N} is evaluated numerically. For reference, we also quote the high-temperature limiting values:
\begin{equation}
\alpha_{\rm MP}^{\rm finite}=\frac{1+\tau}{2},
\qquad
\eta_{\rm MP}^{\rm finite}=\frac{\eta_C}{2},
\label{eq:alpha_mp_finite}
\end{equation}
for the finite near-uniform regime, and
\begin{equation}
\alpha_{\rm MP}^{\rm osc}=\sqrt{\tau},
\qquad
\eta_{\rm MP}^{\rm osc}=1-\sqrt{\tau},
\label{eq:alpha_mp_osc}
\end{equation}
for the oscillator high-temperature limit. We also evaluate two external reference prescriptions.
The finite ecological prescription maximizes a trade-off objective between power and entropy production (commonly $\dot{W}\eta/\eta_C$ or similar) and yields an efficiency of $3\eta_C/4$ in the high-temperature limit.
\begin{equation}
\eta_E^{\rm finite}=\frac{3\eta_C}{4},
\qquad
\alpha_E^{\rm finite}=\frac{1+3\tau}{4},
\label{eq:alpha_ecological}
\end{equation}
and, for the high-temperature harmonic oscillator at the maximum-$\Omega$ operating point~\cite{Singh2022Unified},
\begin{equation}
\eta_\Omega^{\rm osc}
=
1-
\sqrt{\frac{(1-\eta_C)(2-\eta_C)}{2}},
\qquad
\alpha_\Omega^{\rm osc}
=
\sqrt{\frac{\tau(1+\tau)}{2}}.
\label{eq:alpha_omega}
\end{equation}
The finite ecological and oscillator $\Omega$ prescriptions are fixed-cycle-time high-temperature references obtained from trade-off objectives rather than re-optimization of $\mathcal R_N$. Only $\alpha_{\rm MP}^{(N)}$ is re-optimized for each ladder dimension; the remaining prescriptions are external reference points at which Eq.~\eqref{eq:RN_alpha} is evaluated.

\begin{table}[t]
\centering
\caption{
Operating prescriptions used in Fig.~\ref{fig:optimized_reliability}. Only the finite-$N$ maximum-output point $\alpha_{\rm MP}^{(N)}$ is re-optimized for each ladder dimension. The remaining rows are external high-temperature reference prescriptions at which the finite-$N$ reliability is evaluated.
}
\label{tab:operating_points}
\setlength{\tabcolsep}{4pt}
\renewcommand{\arraystretch}{1.2}
\resizebox{\linewidth}{!}{%
\begin{tabular}{l c c}
\hline\hline
Prescription & $\alpha$ & $\eta=1-\alpha$ \\
\hline
finite-$N$ max. output & $\alpha_{\rm MP}^{(N)}$ & $1-\alpha_{\rm MP}^{(N)}$ \\
finite high-$T$ max. output ref. & $(1+\tau)/2$ & $\eta_C/2$ \\
osc. high-$T$ max. output ref. & $\sqrt{\tau}$ & $1-\sqrt{\tau}$ \\
finite ecological ref. & $(1+3\tau)/4$ & $3\eta_C/4$ \\
osc. $\Omega$ ref. & $\sqrt{\tau(1+\tau)/2}$ & $1-\sqrt{\tau(1+\tau)/2}$ \\
\hline\hline
\end{tabular}%
}
\end{table}

\begin{figure}[!htbp]
\centering
\includegraphics[width=\linewidth]{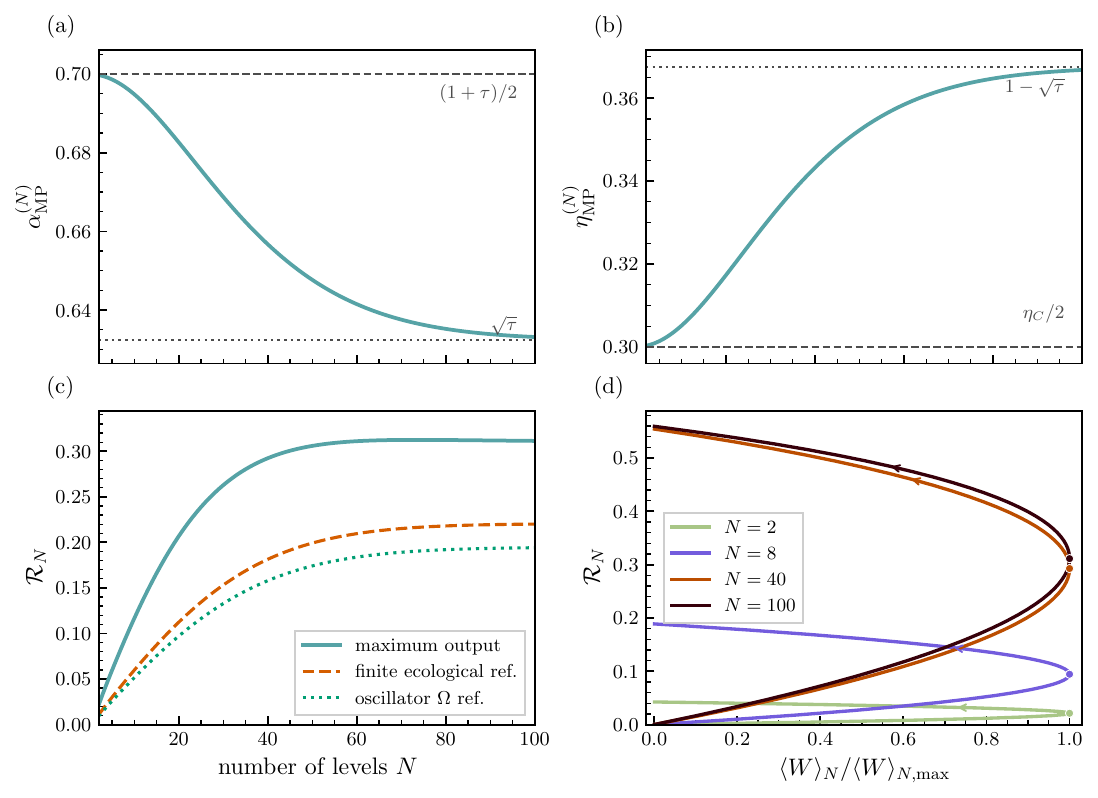}
\caption{
Mean-output optimization and work reliability for a finite uniform ladder at $\tau=T_l/T_h=0.4$ and $z_h=0.08$.
(a) Maximum-output compression ratio $\alpha_{\rm MP}^{(N)}$ obtained from Eq.~\eqref{eq:alpha_mp_N}. Horizontal reference lines show the finite-$N$ high-temperature value $(1+\tau)/2$ and the oscillator high-temperature value $\sqrt{\tau}$.
(b) Efficiency at the same operating point, $\eta_{\rm MP}^{(N)}=1-\alpha_{\rm MP}^{(N)}$, compared with $\eta_C/2$ and $1-\sqrt{\tau}$.
(c) Work reliability $\mathcal R_N$ evaluated at the finite-$N$ maximum-output point and at two external reference prescriptions: the finite high-temperature ecological point and the oscillator $\Omega$ point.
(d) Output--reliability sweep obtained by varying $\alpha$ at fixed $N$. The horizontal axis shows the normalized mean output, $\langle W\rangle_N/\langle W\rangle_{N,\max}$, equivalent to normalized power for fixed cycle time. Filled circles mark the maximum-output points. High reliability away from the maximum-output point should not be interpreted as superior engine performance by itself: in the homothetic benchmark the common work scale $(1-\alpha)\epsilon_h$ cancels from $\mathcal R_N$, so large relative reliability can occur near low-output regions.
}
\label{fig:optimized_reliability}
\end{figure}

Figure~\ref{fig:optimized_reliability} gives the finite-$N$ behavior for $\tau=0.4$ and $z_h=0.08$. Panels (a) and (b) show the crossover of the maximum-output point from the finite high-temperature reference towards the oscillator high-temperature reference as $N$ increases. Panel (c) gives the work reliability at the finite-$N$ maximum-output point and at two external reference prescriptions. Panel (d) plots the output--reliability curve obtained by varying $\alpha$ at fixed $N$; the maximum-output point is generally not the point of largest work reliability.

Thus, optimizing mean output picks a point on the output--reliability curve but does not guarantee minimal relative fluctuations. This is consistent with maximum-power, ecological, and $\Omega$ prescriptions, which optimize mean thermodynamic objectives rather than the relative fluctuations of the single-cycle work output. The reliability ratio we use is a single-cycle, finite-system diagnostic; it is distinct from thermodynamic uncertainty relations, which bound precision via dissipation or entropy production under additional assumptions~\cite{BaratoSeifert2015,Gingrich2016,Pietzonka2018,SaryalAgarwalla2021}.

\section{Incomplete isochores as diagonal athermality}
\label{sec:incomplete_isochores}

The complete-thermalization benchmark assumes that both isochores prepare Gibbs populations independently. To separate incomplete diagonal relaxation from finite-time unitary effects, the strokes remain quasistatic and homothetic, while each isochore is replaced by the diagonal partial-thermalization channel
\begin{equation}
R_s(a|b)=(1-\lambda_s)\delta_{ab}+\lambda_s p_a^s,
\qquad
0\leq \lambda_s\leq 1,
\qquad
s=h,l .
\label{eq:partial_reset_map}
\end{equation}
Here $\lambda_s$ is the reset strength: $\lambda_s=1$ gives complete Gibbs reset, while $\lambda_s=0$ leaves the incoming population unchanged. Equation~\eqref{eq:partial_reset_map} is a phenomenological diagonal channel, not a microscopic finite-time thermalization law; the symbols $\lambda_s$ used here are independent of any parameters appearing in the finite‑time sections. For a specified bath model, the effective relaxation would be derived from the corresponding rate equation or master equation and could depend on the level spacings and transition rates.

In the stationary cold-to-cold cycle, the endpoint populations immediately after the hot and cold isochores satisfy
\[
q^h=R_h q^l,
\qquad
q^l=R_l q^h .
\]
For
\begin{equation}
D_\lambda=\lambda_h+\lambda_l-\lambda_h\lambda_l>0,
\label{eq:Dlambda}
\end{equation}
the stationary solution is unique:
\begin{equation}
q^h
=
\frac{\lambda_h p^h+(1-\lambda_h)\lambda_l p^l}
{D_\lambda},
\qquad
q^l
=
\frac{\lambda_l p^l+(1-\lambda_l)\lambda_h p^h}
{D_\lambda}.
\label{eq:q_partial_main}
\end{equation}
When $\lambda_h=\lambda_l=0$ both isochores act as the identity and any diagonal population with $q^h=q^l$ is stationary; this singular corner is excluded from the closed formulas. Appendix~\ref{app:incomplete_details} gives the full stationary-cycle construction.

For a uniform ladder, the work index in one cold-to-cold cycle is $\Delta=j-i$, where $i$ is the level after the cold isochore and $j$ the level after the hot isochore. The stationary work-index distribution is
\begin{equation}
P_\lambda(\Delta)
=
\sum_i q_i^l R_h(i+\Delta\mid i),
\label{eq:P_lambda}
\end{equation}
with terms outside the allowed level range omitted. The mean work is
\begin{equation}
\langle W\rangle_\lambda
=
(1-\alpha)\epsilon
\frac{\lambda_h\lambda_l}{D_\lambda}
[\nu_N(z_h)-\nu_N(z_l)].
\label{eq:mean_W_partial}
\end{equation}
The variance and reliability follow from the same finite distribution $P_\lambda(\Delta)$; explicit closed forms are given in Appendix~\ref{app:incomplete_details}. The deviation from Gibbs reset is measured by the diagonal athermality
\begin{equation}
\mathcal A_{\rm diag}
=
D_{\rm KL}(q^h\Vert p^h)+D_{\rm KL}(q^l\Vert p^l),
\label{eq:diag_athermality}
\end{equation}
where
\[
D_{\rm KL}(q\Vert p)=\sum_n q_n\ln(q_n/p_n)
\]
is the relative Kullback--Leibler divergence~\cite{Kullback1951}. This quantity measures the statistical distance between the stationary endpoint populations $q^s$ and their Gibbs references $p^s$.

\begin{figure*}[!t]
\centering
\includegraphics[width=0.90\textwidth]{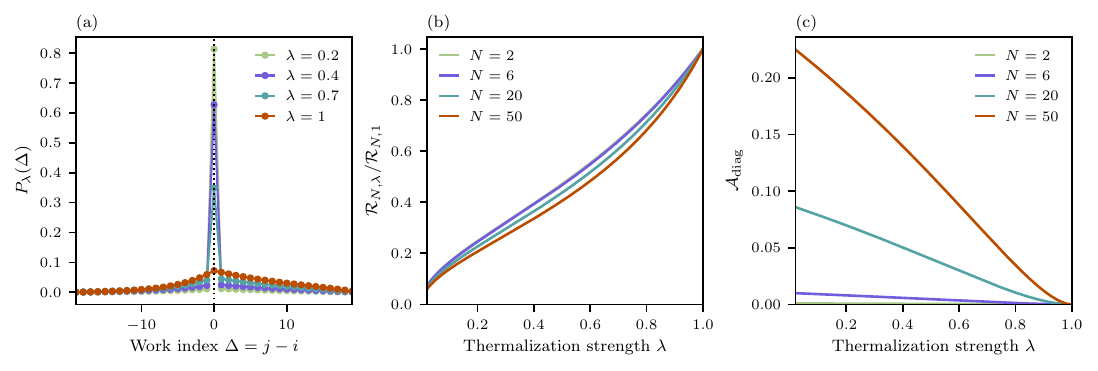}
\caption{
Incomplete isochores as diagonal athermality.
(a) Work-index distribution $P_\lambda(\Delta)$ from Eq.~\eqref{eq:P_lambda} for a uniform ladder at $z_h=0.08$, $z_l=0.20$, and $N=20$, with symmetric thermalization strengths $\lambda_h=\lambda_l=\lambda$. Reducing $\lambda$ transfers probability into the zero-work peak at $\Delta=0$ and suppresses the active nonzero-work part.
(b) Reliability ratio $\mathcal R_{N,\lambda}/\mathcal R_{N,1}$ versus $\lambda$ for representative cutoffs $N$. Incomplete thermalization lowers the work reliability even though the unitary strokes remain quasistatic.
(c) Diagonal athermality $\mathcal A_{\rm diag}=D_{\rm KL}(q^h\Vert p^h)+D_{\rm KL}(q^l\Vert p^l)$ versus $\lambda$. The complete-thermalization limit $\lambda=1$ recovers the Gibbs benchmark.
}
\label{fig:incomplete_isochores}
\end{figure*}

Figure~\ref{fig:incomplete_isochores} shows how partial reset reshapes the work distribution. Decreasing $\lambda$ transfers probability into the zero-work sector and moves the stationary endpoint populations away from their Gibbs references. The quasistatic homothetic efficiency on nonzero-hot-heat trajectories remains fixed at $1-\alpha$, but the extracted work becomes less reliable because the isochores no longer prepare independent Gibbs samples.

\section{Finite-time transition-matrix extensions}
\label{sec:finite_time}

The preceding sections treated the quasistatic homothetic Otto cycle, where each unitary stroke preserves the instantaneous energy label. In that limit a uniform homothetic ladder has trajectory-independent efficiency $1-\alpha$, while the work remains stochastic because the isochores sample thermal occupation numbers. Finite-time strokes modify this structure by allowing transitions between endpoint energy eigenstates.

The transition-matrix identities below apply once the expansion and compression stroke matrices have been specified. They do not determine those matrices from the endpoint spectra alone: the transition probabilities depend on the driving path, stroke duration, nonadiabatic couplings, level structure, and physical implementation. After deriving the general complete-thermalization identities, we evaluate them for a finite-$N$ nearest-neighbor ladder protocol with homothetic endpoints and for the separate harmonic sudden-switch oscillator benchmark. The first is a controlled finite-$N$ stroke model; the second is an oscillator boundary evaluated with numerical cutoffs.

\subsection{General transition-matrix identities}

Complete thermalization on the isochores is retained. The expansion and compression strokes are encoded by
\begin{equation}
T^{\rm e}_{k|n}
=
|\langle k;l|U_{\rm e}|n;h\rangle|^2,
\qquad
T^{\rm c}_{j|m}
=
|\langle j;h|U_{\rm c}|m;l\rangle|^2 .
\label{eq:finite_time_transition_matrices}
\end{equation}
Given $T^{\rm e}$ and $T^{\rm c}$, the following finite sums give the corresponding work moments and, when desired, heat-engine-conditioned efficiency widths. The transition matrices themselves are protocol dependent, so the construction is a transition-matrix framework, not a universal predictive finite-time model.

A finite-time trajectory is $\gamma=(n,k,m,j)$, with probability
\begin{equation}
\Pi_\gamma
=
p_n^h T^{\rm e}_{k|n} p_m^l T^{\rm c}_{j|m},
\label{eq:finite_time_trajectory_weight}
\end{equation}
and work output
\begin{equation}
W_\gamma
=
E_n^h-E_k^l+E_m^l-E_j^h .
\label{eq:finite_time_trajectory_work}
\end{equation}
The corresponding heat variables and conditioned efficiency distribution are given in Appendix~\ref{app:finite_time_details}.

These formulas are evaluated for a single cycle. In the complete-thermalization limit, successive cycles become uncorrelated, and this same distribution directly yields the stationary per-cycle statistics (the extension to $M$ independent cycles is detailed in Appendix~\ref{app:cycle_interpretation}). For a deterministic cycle duration $\tau_{\rm cyc}$, the trajectory power is simply
\begin{equation}
P_\gamma=\frac{W_\gamma}{\tau_{\rm cyc}} .
\label{eq:deterministic_power}
\end{equation}
Consequently,
\begin{equation}
\langle P\rangle=\frac{\langle W\rangle}{\tau_{\rm cyc}},
\qquad
\sigma_P=\frac{\sigma_W}{\tau_{\rm cyc}},
\qquad
\mathcal R_P=\frac{\langle P\rangle}{\sigma_P}
=
\frac{\langle W\rangle}{\sigma_W}
=
\mathcal R_W .
\label{eq:power_work_rescaling}
\end{equation}
Power therefore contains no additional stochastic information beyond work for a fixed cycle time. Stochastic timing, correlations between timing and work, or cycle-to-cycle control noise would require a joint distribution of work and duration.

The raw stochastic efficiency can be singular on trajectories with vanishing absorbed hot heat. Work reliability is therefore used as the main diagnostic, while conditioned efficiency widths are retained only as supplementary trajectory-resolved quantities~\cite{Fei2022,Anka2024}.

For a uniform homothetic ladder,
\begin{equation}
E_n^h=n\epsilon_h,
\qquad
E_n^l=\alpha n\epsilon_h,
\label{eq:finite_time_uniform_ladder}
\end{equation}
finite-time transitions admit a useful jump decomposition. Define
\begin{equation}
d_{\rm e}=k-n,
\qquad
d_{\rm c}=j-m .
\label{eq:finite_time_level_jumps}
\end{equation}
Then
\begin{equation}
\frac{W_\gamma}{\epsilon_h}
=
(1-\alpha)(n-m)-\alpha d_{\rm e}-d_{\rm c}.
\label{eq:finite_time_jump_decomposition}
\end{equation}
The first term is the quasistatic homothetic contribution; the remaining terms are transition-induced jump corrections. Averaging gives
\begin{align}
\frac{\langle W\rangle}{\epsilon_h}
&=
(1-\alpha)(\nu_h-\nu_l)
-
D_{\rm jump},
\nonumber\\
D_{\rm jump}
&=
\alpha\langle d_{\rm e}\rangle_h
+
\langle d_{\rm c}\rangle_l .
\label{eq:finite_time_drift_penalty}
\end{align}
The direct jump-spread contribution to the work variance is
\begin{equation}
V_{\rm jump}
=
\alpha^2{\rm Var}_h(d_{\rm e})
+
{\rm Var}_l(d_{\rm c}),
\label{eq:finite_time_jump_variance}
\end{equation}
with covariance corrections between the initial thermal level and the jump size. The full variance decomposition is given in Appendix~\ref{app:finite_time_details}.

The adiabatic reference reliability is
\begin{equation}
\mathcal{R}_{W,\rm ad}^{(N)}
=
\frac{\nu_N(z_h)-\nu_N(z_l)}
{\sqrt{v_N(z_h)+v_N(z_l)}}.
\label{eq:finite_time_snr_ad_N}
\end{equation}
Under complete thermalization and for a uniform homothetic ladder, the jump decomposition gives the exact reliability ratio
\begin{equation}
\frac{\mathcal R_W}{\mathcal R_{W,\rm ad}}
=
\frac{
1-D_{\rm jump}/[(1-\alpha)(\nu_h-\nu_l)]
}{
\sqrt{
1+(V_{\rm jump}+V_{\rm cov})/[(1-\alpha)^2(v_h+v_l)]
}
}.
\label{eq:finite_time_reliability_ratio}
\end{equation}
Here $V_{\rm cov}$ is the covariance correction given in Appendix~\ref{app:finite_time_details}. Equation~\eqref{eq:finite_time_reliability_ratio} follows directly from the transition-matrix trajectory distribution and is not a weak-driving approximation; a weak finite-time expansion only enters if the transition matrices are subsequently expanded around the adiabatic limit.

Equation~\eqref{eq:finite_time_reliability_ratio} also explains why comparable transition probabilities can suppress small finite-$N$ engines more strongly than oscillator-like ones. For a qubit at high temperature, the excited-state populations are close to one half for both baths, so the thermal occupation lever arm is small while the Bernoulli variance remains finite. For an oscillator, $\nu_\infty(z)\simeq1/z$ and $v_\infty(z)\simeq1/z^2$, giving
\begin{equation}
\mathcal{R}_{W,\rm ad}^{(\infty)}
\simeq
\frac{1/z_h-1/z_l}
{\sqrt{1/z_h^2+1/z_l^2}}.
\label{eq:finite_time_snr_ad_infty}
\end{equation}
The oscillator has larger absolute work fluctuations, but also a larger work signal relative to those fluctuations.

\subsection{Illustrative finite-\texorpdfstring{$N$}{N} ladder protocol with homothetic endpoints}

\begin{figure*}[t]
\centering
\includegraphics[width=1\linewidth]{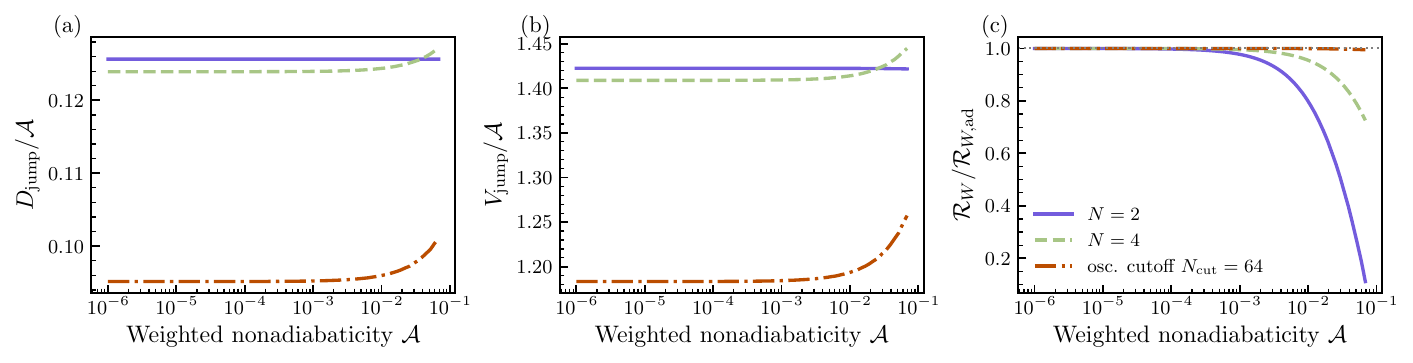}
\caption{
Finite-time transition anatomy for the nearest-neighbor finite-ladder protocol with homothetic endpoints. The parameters are $z_h=0.08$, $z_l=0.20$, $r=z_l/z_h=2.5$, $\alpha=0.65$, and $\tau=T_l/T_h=\alpha z_h/z_l=0.26$. The propagators are computed with $\epsilon_h=\hbar=1$, dimensionless stroke duration $\Theta=\epsilon_h\tau_u/\hbar=1$, and $N_t=70$ midpoint time slices. Results are shown for $N=2$, $N=4$, and a large-cutoff ladder reference with $N_{\rm cut}=64$. The horizontal axis is the weighted nonadiabaticity $\mathcal A$, obtained by sweeping the dimensionless mixing amplitude $g_N/\epsilon_h$ separately for each cutoff and interpolating the resulting data onto a common range.
(a) Transition-induced drift per unit nonadiabaticity, $D_{\rm jump}/\mathcal A$.
(b) Transition-induced spread per unit nonadiabaticity, $V_{\rm jump}/\mathcal A$.
(c) Work-reliability ratio relative to the adiabatic homothetic benchmark, $\mathcal R_W/\mathcal R_{W,\rm ad}$.
The nearly constant behavior in panels (a) and (b) is a feature of this nearest-neighbor protocol, not a universal finite-time Otto result.
}
\label{fig:finite_time_transition_anatomy}
\end{figure*}

A dimension-resolved nearest-neighbor protocol is used as a reproducible finite-$N$ example. The endpoint Hamiltonians remain exactly homothetic, while the stroke contains a noncommuting term that generates finite-time transitions. For dimension $N$, define
\begin{equation}
n_N=\sum_{n=0}^{N-1} n\, |n\rangle\langle n|,
\qquad
a_N=\sum_{n=1}^{N-1} \sqrt n\, |n-1\rangle\langle n|,
\label{eq:finite_time_number_and_ladder}
\end{equation}
and the normalized mixing operator
\begin{equation}
V_N
=
\frac{a_N+a_N^\dagger}
{\|a_N+a_N^\dagger\|_2}.
\label{eq:finite_time_mixing_operator}
\end{equation}
The expansion stroke is generated by
\begin{align}
H_{\rm e}^{(N)}(s)
&=
\epsilon_h[1-(1-\alpha)s]\, n_N
+
g_N\sin^2(\pi s)\, V_N,
\nonumber\\
&0\le s\le1,
\label{eq:finite_time_expansion_protocol}
\end{align}
and the compression stroke by
\begin{equation}
H_{\rm c}^{(N)}(s)
=
\epsilon_h[\alpha+(1-\alpha)s]\, n_N
+
g_N\sin^2(\pi s)\, V_N .
\label{eq:finite_time_compression_protocol}
\end{equation}
The mixing term vanishes at the endpoints, so the measured endpoint Hamiltonians are
\begin{equation}
H_h^{(N)}=\epsilon_h n_N,
\qquad
H_l^{(N)}=\alpha\epsilon_h n_N .
\label{eq:finite_time_endpoint_hamiltonians}
\end{equation}
For Fig.~\ref{fig:finite_time_transition_anatomy}, $\epsilon_h=\hbar=1$ and the expansion and compression strokes use the same dimensionless duration
\begin{equation}
\Theta=\frac{\epsilon_h\tau_u}{\hbar}=1 .
\end{equation}
With $t=\tau_u s$, the time-ordered propagators are computed by a midpoint product formula with $N_t=70$ uniform time slices.

The departure from adiabatic label preservation is measured by the weighted nonadiabaticity
\begin{equation}
\mathcal A
=
\frac12
\left[
1-\sum_n p_n^h T^{\rm e}_{n|n}
+
1-\sum_m p_m^l T^{\rm c}_{m|m}
\right].
\label{eq:finite_time_weighted_A}
\end{equation}
The raw amplitude $g_N$ is not used as the comparison axis because the same value of $g_N$ produces different transition probabilities at different dimensions. Instead, $g_N/\epsilon_h$ is swept separately for each cutoff, and the data are reparametrized by the achieved value of $\mathcal A$. The plotted curves are interpolated onto a common reliability-safe interval where all representative cases satisfy $\mathcal R_W/\mathcal R_{W,\rm ad}>0$.

Figure~\ref{fig:finite_time_transition_anatomy} shows the transition-induced drift and spread per unit weighted nonadiabaticity, $D_{\rm jump}/\mathcal A$ and $V_{\rm jump}/\mathcal A$. Their near constancy over the plotted interval is a property of the chosen nearest-neighbor model. The main point is the separation between the adiabatic finite-$N$ work fluctuations and the additional nonadiabatic drift and jump-spread contributions generated by a specified transition matrix.

\subsection{Harmonic sudden-switch oscillator benchmark}

The nearest-neighbor protocol above is a finite-$N$ ladder model with homothetic endpoints. The harmonic sudden switch is a different boundary case: an oscillator frequency quench, approximated numerically with a cutoff when transition matrices are evaluated.

A sudden gap change in the strict finite uniform ladder is commuting. For $H_h=\epsilon_h n_N$ and $H_l=\alpha\epsilon_h n_N$, the endpoint Hamiltonians share eigenvectors, so an instantaneous gap rescaling gives $T_{k|n}=\delta_{kn}$. A finite uniform ladder has no sudden-switch friction unless the stroke contains a noncommuting ingredient. The harmonic oscillator differs because changing the frequency changes the eigenbasis, so an instantaneous quench $\omega_h\to\omega_l$ produces squeezed-number-state overlaps and genuine transition broadening.

The oscillator quench is commonly organized by the nonadiabaticity parameter $Q^*$~\cite{RezekKosloff2006,SaryalAgarwalla2021,Singh2022Unified}. For $\omega_h\to\omega_l=\alpha\omega_h$,
\begin{equation}
Q^*
=
\frac{\omega_h^2+\omega_l^2}{2\omega_h\omega_l}
=
\frac{1+\alpha^2}{2\alpha}.
\label{eq:finite_time_Qstar}
\end{equation}
The sudden-switch mean work is the adiabatic homothetic value minus a friction penalty,
\begin{equation}
\langle W\rangle_{\rm ss}
=
\langle W\rangle_{\rm ad}
-
W_{\rm fric},
\label{eq:finite_time_sudden_mean}
\end{equation}
where
\begin{equation}
\langle W\rangle_{\rm ad}
=
(1-\alpha)\epsilon_h(\nu_h-\nu_l),
\label{eq:finite_time_sudden_ad_work}
\end{equation}
and
\begin{equation}
W_{\rm fric}
=
\epsilon_h(Q^*-1)
\left[
\alpha\left(\nu_h+\frac12\right)
+
\left(\nu_l+\frac12\right)
\right].
\label{eq:finite_time_Qstar_friction}
\end{equation}
The variance contains a squeezed-transition contribution proportional to $(Q^*)^2-1$, derived in Appendix~\ref{app:finite_time_details}.

\begin{figure*}[t]
\centering
\includegraphics[width=0.85\linewidth]{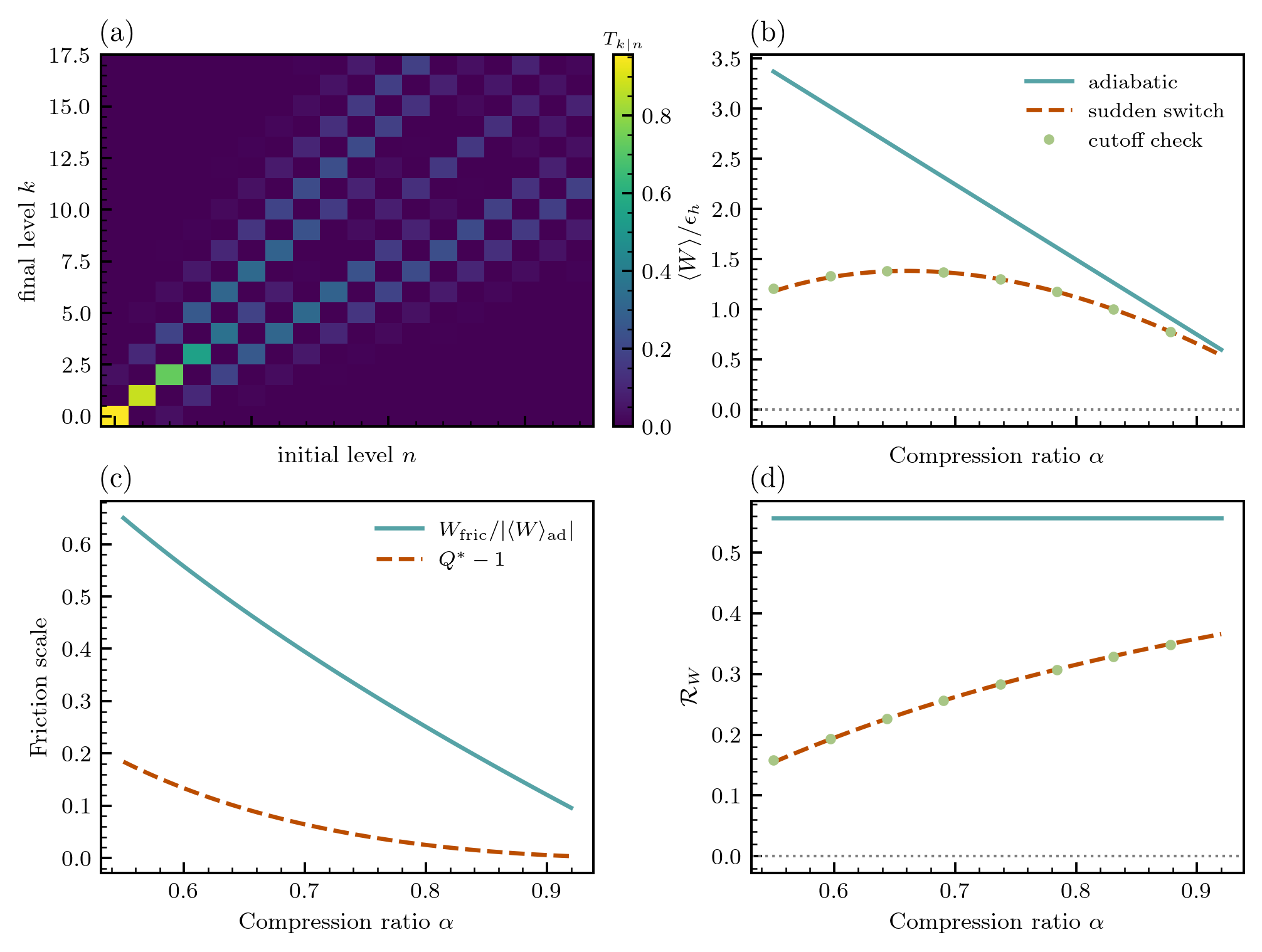}
\caption{
Harmonic sudden-switch oscillator benchmark organized by the nonadiabaticity factor $Q^*$. Here $N_{\rm cut}$ denotes a numerical oscillator cutoff used to approximate the oscillator transition matrix; it should not be interpreted as a physical finite uniform ladder.
(a) Sudden-switch transition matrix $T_{k|n}=|\langle k;\omega_l|n;\omega_h\rangle|^2$ at $\alpha=\omega_l/\omega_h=0.55$, showing the same-parity structure of squeezed-number-state overlaps.
(b) Mean work in the adiabatic and sudden-switch protocols. The sudden switch subtracts the friction penalty $(Q^*-1)[\alpha(\nu_h+1/2)+(\nu_l+1/2)]$, shifting the engine boundary. Sparse markers show direct finite-cutoff TPM evaluations from the truncated transition matrix.
(c) Relative friction scale and $Q^*-1$, with $Q^*=(1+\alpha^2)/(2\alpha)$.
(d) Work reliability in the adiabatic and sudden-switch protocols, again compared with finite-cutoff TPM markers.
The finite-cutoff markers are numerical checks of the oscillator transition-matrix implementation and should be read together with the cutoff-stabilization test in Table~\ref{tab:sudden_switch_stabilization}; small cutoffs are not assumed to be converged.
}
\label{fig:harmonic_Qstar_sudden_switch}
\end{figure*}

Figure~\ref{fig:harmonic_Qstar_sudden_switch} shows the oscillator sudden-switch boundary. The transition matrix in panel (a) displays the same-parity structure of squeezed-number-state overlaps generated by the instantaneous frequency switch. Panels (b) and (d) compare the analytic $Q^*$ expressions with direct finite-cutoff TPM evaluations. The cutoff calculation is a numerical check of the oscillator boundary, not a physical finite-$N$ ladder engine. In finite ladders, reliability loss from a noncommuting stroke is organized by jump drift and jump spread; in the harmonic sudden switch, it is organized by the $Q^*$-controlled friction penalty and by the squeezed-transition contribution to the variance.

An additional noncommuting endpoint stress test is given in Appendix~\ref{app:finite_time_details}. That construction is not part of the exact homothetic finite-time benchmark, because its endpoint spectra are generally not related by common gap rescaling for nonzero mixing.

\section{Weakly nonhomothetic spectra}
\label{sec:nonhom}

Exact homothety removes quasistatic stochastic-efficiency fluctuations. Once the endpoint spectra are no longer related by common gap rescaling, efficiency fluctuations reappear. We parameterize a weak distortion of the low spectrum by
\begin{equation}
E_n^l=\alpha E_n^h+\chi+\delta_n ,
\label{eq:nonhom_spectrum}
\end{equation}
where $\delta_n$ is level dependent. A constant contribution to $\delta_n$ can be absorbed into $\chi$ and has no effect on work differences. The perturbative treatment that follows assumes $|\beta_l\delta_n|\ll1$ for the thermally relevant levels, so that the cold Gibbs weights can be expanded to first order in the distortion.

For a quasistatic trajectory labelled by the hot and cold thermal indices $(n,m)$, the extracted work is
\begin{equation}
W(n,m)=(1-\alpha)(E_n^h-E_m^h)-(\delta_n-\delta_m).
\label{eq:nonhom_work}
\end{equation}
Introduce the level-dependent gap-change variable
\begin{equation}
g_n=E_n^h-E_n^l=(1-\alpha)E_n^h-\delta_n-\chi .
\end{equation}
Since $\chi$ cancels from $g_n-g_m$,
\begin{equation}
W(n,m)=g_n-g_m,
\end{equation}
and therefore
\begin{equation}
\langle W\rangle_N=\langle g\rangle_h-\langle g\rangle_l,
\qquad
\sigma_{W,N}^2=\Var_h(g)+\Var_l(g).
\label{eq:nonhom_exact_g}
\end{equation}
Equivalently,
\begin{align}
\langle W\rangle_N
&=(1-\alpha)[\langle E^h\rangle_h-\langle E^h\rangle_l]
-[\langle\delta\rangle_h-\langle\delta\rangle_l],
\label{eq:nonhom_mean_exact}\\
\sigma_{W,N}^2
&=(1-\alpha)^2[\Var_h(E^h)+\Var_l(E^h)]
\nonumber\\
&\quad +[\Var_h(\delta)+\Var_l(\delta)]
\nonumber\\
&\quad -2(1-\alpha)[\Cov_h(E^h,\delta)+\Cov_l(E^h,\delta)] .
\label{eq:nonhom_var_exact}
\end{align}
The moment-reduction structure survives, but the work moments are no longer determined solely by the hot-spectrum moments. Additional contributions arise from the distorted-gap statistics and their covariance with the thermal populations.

For trajectories with $E_n^h\ne E_m^h$,
\begin{equation}
\eta_{\st}(n,m)
=
1-\alpha
-
\frac{\delta_n-\delta_m}{E_n^h-E_m^h}.
\label{eq:eta_nonhom}
\end{equation}
Level-dependent nonhomothety therefore reintroduces quasistatic efficiency fluctuations. In a finite spectrum, the nominal parameter $\alpha$ is not sufficient to characterize the cycle unless all populated gaps scale by the same factor. For adjacent gaps one may define
\begin{equation}
\alpha_q
=
\frac{E_{q+1}^l-E_q^l}
{E_{q+1}^h-E_q^h}.
\label{eq:nonhom_gap_ratios}
\end{equation}
If all $\alpha_q$ are equal over the thermally populated part of the spectrum, the endpoint spectra are effectively homothetic. If the $\alpha_q$ vary, different trajectories sample different effective compression ratios, so both the work distribution and the stochastic-efficiency distribution change even when the same nominal $\alpha$ is used. The sign and magnitude of the correction depend on the distorted gaps that are thermally occupied and on their covariance with the endpoint Gibbs weights.

The stochastic-efficiency diagnostics are evaluated on the heat-engine trajectory set
\begin{equation}
\Gamma_{\rm HE}
=
\{(n,m):Q_h(n,m)>0,\ Q_l(n,m)<0,\ W(n,m)>0\},
\label{eq:nonhom_HE_set}
\end{equation}
where
\begin{equation}
Q_h(n,m)=E_n^h-E_m^h,
\qquad
Q_l(n,m)=E_m^l-E_n^l .
\end{equation}
The corresponding trajectory weight is
\begin{equation}
\mathcal N_{\rm HE}
=
\sum_{(n,m)\in\Gamma_{\rm HE}}p_n^h p_m^l .
\label{eq:nonhom_HE_weight}
\end{equation}
Conditioned efficiency moments are
\begin{equation}
\langle \eta^r\rangle_{\rm HE}
=
\frac{1}{\mathcal N_{\rm HE}}
\sum_{(n,m)\in\Gamma_{\rm HE}}
p_n^h p_m^l
\left[
\frac{W(n,m)}{Q_h(n,m)}
\right]^r ,
\label{eq:nonhom_eta_moments_HE}
\end{equation}
and the plotted efficiency width is
\begin{equation}
\sigma_\eta
=
\sqrt{
\langle \eta^2\rangle_{\rm HE}
-
\langle \eta\rangle_{\rm HE}^2
}.
\label{eq:nonhom_sigma_eta}
\end{equation}
The conditioning removes the singular $Q_h=0$ trajectories and restricts the statistics to genuine engine cycles.

Expand the cold Gibbs weights around the homothetic reference distribution,
\[
p_n^{l,0}\propto\exp[-\beta_l\alpha E_n^h].
\]
To first order in $\delta_n$,
\begin{equation}
p_n^l
=
p_n^{l,0}
\left[
1-\beta_l(\delta_n-\langle\delta\rangle_l^{(0)})
\right]
+O(\delta^2).
\end{equation}
The first-order mean-work correction is
\begin{equation}
\delta\langle W\rangle^{(1)}
=
-[\langle\delta\rangle_h-\langle\delta\rangle_l^{(0)}]
+
(1-\alpha)\beta_l\Cov_l^{(0)}(E^h,\delta).
\label{eq:mean_corr}
\end{equation}
The first-order variance correction is
\begin{align}
\delta\sigma_W^{2(1)}
&=
-2(1-\alpha)
[
\Cov_h(E^h,\delta)
+
\Cov_l^{(0)}(E^h,\delta)
]
\nonumber\\
&\quad
-(1-\alpha)^2\beta_l
\Cov_l^{(0)}((E^h-\mu_l^{(0)})^2,\delta).
\label{eq:var_corr}
\end{align}
Consequently,
\begin{equation}
\delta\mathcal{R}^{(1)}
=
\frac{\delta\langle W\rangle^{(1)}}{\sigma_{W,0}}
-
\frac{\langle W\rangle_0}{2\sigma_{W,0}^3}
\delta\sigma_W^{2(1)}.
\label{eq:nonhom_reliability_corr}
\end{equation}

\begin{figure*}[!htbp]
\centering
\includegraphics[width=1\linewidth]{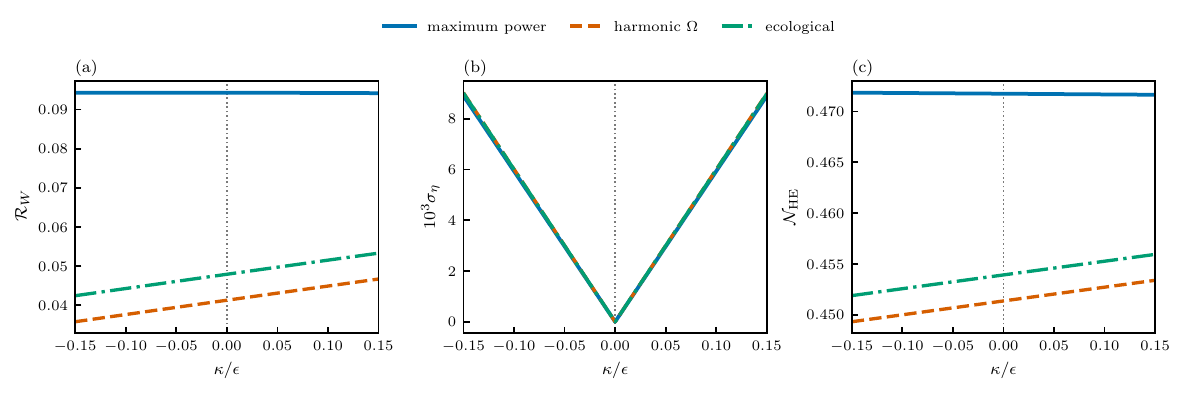}
\caption{
Weak nonhomothetic distortion of the finite ladder,
$E_n^l=\alpha E_n^h+\kappa\epsilon[n/(N-1)]^2$, with $N=12$, $z_h=0.10$, and $\tau=T_l/T_h=0.4$. The nominal compression ratio $\alpha$ is fixed by the corresponding homothetic operating prescription before the distortion is applied. The three curves show the maximum-output prescription, the harmonic $\Omega$ reference prescription, and the ecological reference prescription.
(a) Work reliability $\mathcal R_W$ computed from the exact distorted-spectrum finite sums.
(b) Heat-engine-conditioned stochastic-efficiency width $\sigma_\eta$, plotted as $10^3\sigma_\eta$.
(c) Heat-engine trajectory weight $\mathcal N_{\rm HE}$.
The vertical dotted line marks the homothetic point $\kappa=0$, where quasistatic stochastic-efficiency fluctuations vanish.
}
\label{fig:nonhom_optimized}
\end{figure*}
\begin{table*}[t]
\centering
\caption{Summary of key work moments and reliability expressions for the homothetic Otto engine and its extensions. The common work scale is $\Delta W_1 = (1-\alpha)\epsilon$ (uniform ladder). Abbreviations: ``hom.'' = homothetic, ``HT'' = high temperature, ``osc.'' = oscillator limit. The finite-time results refer to a complete thermalization cycle with transition matrices $T^{\rm e},T^{\rm c}$; $d_{\rm e},d_{\rm c}$ are the level jumps, and $V_{\rm cov}$ is the covariance correction (Appendix~\ref{app:finite_time_details}). Incomplete‑isochore formulas use the partial‑thermalization strengths $\lambda_h,\lambda_l$ and $D_\lambda=\lambda_h+\lambda_l-\lambda_h\lambda_l$.}
\label{tab:summary_formulas}
\begin{ruledtabular}
\begin{tabular}{p{3.2cm}p{4.5cm}p{4.5cm}p{3.5cm}}
\hline\hline
\textbf{Scenario} & \textbf{Mean work $\langle W\rangle_N$} & \textbf{Variance $\sigma_{W,N}^2$} & \textbf{Reliability $\mathcal R_N$ (or ratio)} \\
\hline
complete thermalization, arbitrary hom. spectrum & $(1-\alpha)(\mu_h-\mu_l^{(h)})$ & $(1-\alpha)^2[\Var_h(E^h)+\Var_l(E^h)]$ & $\displaystyle\frac{\mu_h-\mu_l^{(h)}}{\sqrt{\Var_h(E^h)+\Var_l(E^h)}}$ \\
[0.5ex]
Uniform ladder (exact) & $\Delta W_1\,[\nu_N(z_h)-\nu_N(z_l)]$ & $(\Delta W_1)^2\,[v_N(z_h)+v_N(z_l)]$ & $\displaystyle\frac{\nu_N(z_h)-\nu_N(z_l)}{\sqrt{v_N(z_h)+v_N(z_l)}}$ \\
[0.5ex]
Qubit $(N=2)$ & $\Delta W_1\,(p_h-p_l)$ & $(\Delta W_1)^2[p_h(1-p_h)+p_l(1-p_l)]$ & $\displaystyle\frac{p_h-p_l}{\sqrt{p_h(1-p_h)+p_l(1-p_l)}}$ \\
[0.5ex]
Oscillator $(N\to\infty)$ & $\Delta W_1\bigl(\frac{1}{e^{z_h}-1}-\frac{1}{e^{z_l}-1}\bigr)$ & $(\Delta W_1)^2\bigl[\frac{e^{z_h}}{(e^{z_h}-1)^2}+\frac{e^{z_l}}{(e^{z_l}-1)^2}\bigr]$ & Eq.~\eqref{eq:snr_inf} \\
[0.5ex]
Near-uniform HT $(N z_h,N z_l\ll1)$ & $\Delta W_1\frac{(z_l-z_h)(N^2-1)}{12}$ & $(\Delta W_1)^2\frac{N^2-1}{12}$ & $\displaystyle (z_l-z_h)\sqrt{\frac{N^2-1}{24}}$ \\
[0.5ex]
HT osc. plateau $(z_h\to0,\ r=z_l/z_h)$ & — & — & $\displaystyle\frac{r-1}{\sqrt{1+r^2}}$ \\
\hline
Incomplete isochores (diagonal) & $\displaystyle\frac{\lambda_h\lambda_l}{D_\lambda}\langle W\rangle_N^{\rm reset}$ & See Appendix~\ref{app:incomplete_details} & $\mathcal R_{N,\lambda}<\mathcal R_{N,1}$ (Fig.~\ref{fig:incomplete_isochores}) \\
\hline
Finite-time, uniform ladder & $\langle W\rangle_{\rm ad}-\epsilon_h D_{\rm jump}$ & $(\Delta W_1)^2(v_h+v_l)+\epsilon_h^2(V_{\rm jump}+V_{\rm cov})$ & $\displaystyle\frac{\mathcal R_W}{\mathcal R_{W,\rm ad}} = \frac{1-\frac{D_{\rm jump}}{(1-\alpha)(\nu_h-\nu_l)}}{\sqrt{1+\frac{V_{\rm jump}+V_{\rm cov}}{(1-\alpha)^2(v_h+v_l)}}}$ \\
[0.5ex]
Harmonic sudden switch & $\langle W\rangle_{\rm ad} - W_{\rm fric}$ (Eq.~\eqref{eq:finite_time_Qstar_friction}) & Eq.~\eqref{eq:app_ft_sudden_variance} & $\mathcal R_{\rm ss}<\mathcal R_{\rm ad}$ (Fig.~\ref{fig:harmonic_Qstar_sudden_switch}) \\
\hline
Weakly nonhom. (first order) & $\langle W\rangle_0 + \delta\langle W\rangle^{(1)}$ & $\sigma_{W,0}^2 + \delta\sigma_W^{2(1)}$ & $\mathcal R_0 + \delta\mathcal R^{(1)}$ (Eqs.~\eqref{eq:mean_corr}--\eqref{eq:nonhom_reliability_corr}) \\
\hline\hline
\end{tabular}
\end{ruledtabular}
\end{table*}

Figure~\ref{fig:nonhom_optimized} illustrates the response of the exact finite-spectrum diagnostics to a quadratic gap distortion. At $\kappa=0$ the spectra are homothetic and the conditioned quasistatic efficiency width vanishes. Away from this point, different trajectories experience different effective compression ratios, producing a finite $\sigma_\eta$. The work reliability and the heat-engine trajectory weight also change, with the magnitude and direction of the correction determined by the distorted gaps that are thermally populated and by their covariance with the endpoint Gibbs weights.

Anharmonic traps, interacting spectra, avoided crossings, critical rearrangements, and dressed light--matter spectra can all violate exact common gap rescaling. Existing studies of nonharmonic traps, multigap constraints, critical engines, and Rabi--Stark engines show that spectral nonuniformity can modify work statistics, engine feasibility, and mean performance~\cite{ZhengPoletti2015,SonkarJohal2025,CampisiFazio2016,HolubecRyabov2017,Mukherjee2024CriticalFiniteTemp,Xu2024AQRSM}. The perturbative expansion used here is limited to weak spectral distortions where $|\beta_l\delta_n|\ll1$. Strong dressing, parity-sector restructuring, avoided crossings, or gap inversions lie outside its range of validity and require the exact finite-spectrum sums.

\section{Physical interpretation and platforms}
\label{sec:platforms}

The homothetic benchmark separates the Otto efficiency from the reliability of the work output. Qubits, finite qudits, and oscillators can share the same quasistatic Otto efficiency while producing different TPM work distributions. Different platforms therefore probe different parts of the benchmark.

Spin and nuclear-magnetic-resonance implementations are closest to the qubit boundary. They can test the two-level reduction, finite-time work statistics, and reliability near optimized operation. The NMR experiment of Peterson \textit{et al.} is especially relevant because it reconstructs work and heat statistics in a spin Otto engine~\cite{Peterson2019}. Trapped-ion motional modes and single-atom heat engines are closer to the oscillator boundary and to finite-time harmonic protocols, including sudden-switch and shortcut-assisted frequency modulation~\cite{Abah2012,Rossnagel2016,DelCampo2014}. A spin engine coupled to a harmonic flywheel gives direct access to stored-work fluctuations and reliability diagnostics~\cite{Lindenfels2019}.

Finite ladders and weakly anharmonic qudits occur naturally in superconducting circuits and engineered synthetic spectra. Transmons, flux-tunable circuits, and resonators can interpolate between qubit-like, finite-ladder, and oscillator-like regimes. In these intermediate finite‑$N$ platforms, the finite‑size truncation contributions to the Massieu potential of the working medium dictate the thermal fluctuations~\cite{Hoyuelos2025}, making it explicit how a finite heat capacity intrinsically limits the reliability of work extraction. Recent dissipation-engineered superconducting-circuit heat-engine experiments make this platform a plausible setting for testing finite‑$N$ reliability and nonhomothetic corrections~\cite{Marchegiani2016,Uusnakki2026}. Collisional reservoirs and ultracold atomic systems provide complementary routes to tunable multilevel working media~\cite{Bouton2021}. These platforms do not realize all assumptions of the benchmark simultaneously; rather, they isolate specific regimes or failure modes, such as finite spectral support, nonadiabatic transitions, incomplete thermalization, or spectral non-self-similarity.

\subsection*{When does a finite ladder become oscillator-like? — a practical guideline}

The dimensionless product $N z_h$ controls the crossover. For the uniform ladder, $N z_h \simeq \beta_h E_{\max}$, where $E_{\max}=N\epsilon$ is the upper edge of the spectrum. When $N z_h \gg 1$, the thermally populated tail is fully resolved before the boundary is felt; the finite ladder is effectively cutoff‑insensitive and closely mimics the oscillator. When $N z_h \lesssim 1$, the whole ladder is thermally populated, finite‑size effects dominate, and the engine operates in a regime that interpolates between the qubit and the oscillator.

A concrete illustration can be drawn from the one‑percent design maps in Appendix~\ref{app:dimension_design} (summarized in Fig.~\ref{fig:asymptotic_saturation}). For $z_h=0.08$ and a ratio $r=z_l/z_h=2.5$, a ladder with $N\simeq10^2$ is needed to reproduce the oscillator reliability to within $1\%$. However, much of the reliability gain over a qubit is already captured by far smaller ladders: Fig.~\ref{fig:snr_crossover}(a) shows that for $z_h=0.08$ an $N=6$ ladder already reaches roughly $60$–$70\%$ of the oscillator reliability, and $N=30$ is within a few percent. Thus a finite qudit with $N\sim 10$–$30$ serves as a compact testbed for finite‑size fluctuation physics, while requiring only a modest spectral span. Conversely, if one wishes to use a harmonic‑oscillator model to describe an actual finite ladder, the condition $N z_h \gg 1$ should be checked; when it fails, the finite‑size corrections derived in this paper must be included.

Recent results in quantum metrology further reinforce this spectral dichotomy: finite‑spectrum probes and unbounded continuous oscillators belong to different high‑temperature universality classes, leading to distinct bounds on the quantum Fisher information~\cite{Aiache2026}. Thus the choice of $N$ is not merely a technical detail but can qualitatively affect the predicted sensitivity and fluctuations.

\subsection*{Experimental parameter mapping}

As a scale estimate, consider a microwave finite ladder or resonator with hot frequency $f_h=5\,{\rm GHz}$, so that $hf_h/k_B\simeq 240\,{\rm mK}$. The dimensionless hot parameter is
\begin{equation}
z_h=\frac{hf_h}{k_B T_h}.
\end{equation}
The representative value $z_h=0.08$ used in several figures corresponds to an engineered hot temperature $T_h\simeq 3.0\,{\rm K}$. If the low frequency is $f_l=\alpha f_h$ with $\alpha=0.55$, then $f_l=2.75\,{\rm GHz}$. The value $z_l=0.20$ corresponds to $T_l\simeq 0.66\,{\rm K}$, giving $r=z_l/z_h=2.5$ and $\tau=T_l/T_h\simeq0.22$, within the positive-work regime $\alpha>\tau$. The work-index spacing is
\begin{equation}
\Delta W=(1-\alpha)hf_h,
\end{equation}
which equals $h\times2.25\,{\rm GHz}$, or about $108\,{\rm mK}$ in $k_B$ units, for these parameters. In the quasistatic homothetic benchmark, repeated endpoint level measurements would reconstruct the integer work-index distribution $k=n-m$, with $W_k=(1-\alpha)hf_h k$.

The reset strength and nonadiabaticity parameters can also be calibrated experimentally. If a diagonal relaxation stage is approximately single-rate, then
\begin{equation}
\lambda_s=1-\exp(-\Gamma_s t_s).
\end{equation}
Thus $\lambda_s=0.90$ corresponds to $\Gamma_s t_s\simeq2.3$, while $\lambda_s=0.99$ corresponds to $\Gamma_s t_s\simeq4.6$. This identification is model dependent, but it links the partial-reset parameter to population-relaxation data. Similarly, the weighted nonadiabaticity $\mathcal A$ is obtained from the measured transition matrix. If the thermally weighted same-level probability is $0.995$ on both strokes, then $\mathcal A\simeq5\times10^{-3}$, placing the operation in the weak-transition regime of the finite-time diagnostic. The dimensionless variables in the theory can therefore be mapped to reported frequencies, temperatures, level cutoffs, relaxation strengths, transition matrices, and measured work histograms.

The formulas also give a baseline for diagonal nonequilibrium reservoirs. If the isochores prepare diagonal but non-Gibbs steady states $q_n^h$ and $q_m^l$, as may occur for engineered nonequilibrium reservoirs after dephasing, the quasistatic work formulas remain valid after replacing $p^h,p^l$ by those endpoint distributions. The operating condition, optimized points, and reliability are then controlled by the nonequilibrium populations rather than by the thermal parameters $z_h$ and $z_l$.

If the reservoirs create or preserve energy-basis coherences, this replacement is no longer sufficient. Coherent resources, squeezed reservoirs, and other nonthermal baths can modify entropy flow, apparent efficiency bounds, power fluctuations, and the interpretation of heat and work~\cite{Manzano2016SqueezedReservoir,Klaers2017SqueezedReservoir,Niedenzu2018BeyondSecondLaw,Xiao2023SqueezedFluctuations}. In that regime, TPM statistics describe a dephased, measurement-conditioned engine. DBN or related minimally invasive formulations are required to track coherent contributions to work and heat. Comparing TPM and DBN statistics for finite-$N$ homothetic engines with coherent or squeezed reservoirs remains outside the diagonal-Gibbs setting treated here.
\section{Summary of main formulas}
\label{sec:summary_table}

For quick reference, Table~\ref{tab:summary_formulas} collects the central expressions derived in this work. All quantities refer to a finite $N$-level working medium; $z_h=\beta_h\epsilon$, $z_l=\beta_l\alpha\epsilon$, and $\nu_N$, $v_N$ are defined in Eqs.~\eqref{eq:nuN}--\eqref{eq:vN}.

\section{Conclusion}
\label{sec:conclusion}

We derived exact finite‑$N$ two‑point‑measurement work statistics for homothetic quantum Otto engines. Common gap rescaling freezes the quasistatic stochastic efficiency, isolating work fluctuations as a pure probe of finite Hilbert‑space support. For a uniform ladder we obtained closed‑form expressions for the full work distribution, its moments, its cumulant‑generating function, and the signal‑to‑width reliability. These formulas connect the qubit and oscillator limits, capture a low‑temperature effective two‑level regime, and reveal a striking non‑commutation of the high‑temperature and infinite‑dimensional limits.

This non‑commutation is a central result. At fixed finite $N$, the high‑temperature endpoint is the maximally mixed state on a bounded support; the hot and cold Gibbs states become indistinguishable and work reliability vanishes. In contrast, the oscillator has no normalizable infinite‑temperature Gibbs state, and the high‑temperature limit instead follows an ever‑expanding thermal tail, leaving a finite reliability plateau. Consequently, modeling a highly excited finite‑dimensional engine as a continuous harmonic oscillator can drastically overestimate its thermodynamic reliability by overlooking the saturation imposed by a finite spectrum. The deep consequences of this bounded‑versus‑unbounded dichotomy for general thermodynamic uncertainty relations, isothermal cycles, and autonomous machines will be explored in a separate work.

The finite‑$N$ benchmark also yields a practical useful‑dimension criterion: once the thermally populated part of the spectrum is resolved, further enlarging the Hilbert space gives diminishing returns for work reliability. We showed that mean‑output optimization and work constancy are distinct objectives, and that standard maximum‑power, ecological (power–efficiency trade‑off), or $\Omega$‑type prescriptions select operating points with very different relative fluctuations.

Extending the benchmark to incomplete isochores, finite‑time strokes, and weak spectral non‑homothety, we isolated diagonal athermality, transition‑induced drift and jump spread, and level‑dependent gap distortions as separate, additive contributions to work unreliability. The entire framework uses only diagonal endpoint preparations and TPM statistics, providing a clean thermal reference against which coherent, squeezed‑reservoir, or non‑passive effects can be compared.

In summary, this work delivers a complete, analytically tractable null model for diagnosing and disentangling the sources of work unreliability in finite‑dimensional quantum Otto engines, and it establishes a sharp boundary between the finite and the infinite in quantum thermodynamic fluctuations.

\begin{acknowledgments}
This study was financed in part by the Coordenação de Aperfeiçoamento de Pessoal de Nível Superior - Brasil (CAPES) - Finance Code 001, and by the National Natural Science Foundation of China (NSFC) under Grant No. 12174346. C.C acknowledges the  support from Fundação de Amparo à Pesquisa do Estado da Bahia - FAPESB under grant numbers BOL2809/2025 and PPP0006/2024.  N.G.A. acknowledges support from FAPESP under Grant No. 2024/21707-0.
\end{acknowledgments}

\section*{Data and code availability}

The data underlying the figures and the numerical scripts used to generate them are available from the corresponding author upon reasonable request. All analytic formulas needed to reproduce the main finite-ladder results are given in the text and appendices.

\subsection*{Artificial Intelligence Usage Declaration}
In accordance with journal guidelines, the authors declare the use of Artificial Intelligence Generated Content (AIGC) tools during the preparation of this manuscript. Specifically, Gemini/ChatGPT was utilized for language polishing, text editing, and assisting with deep literature research. Following the use of these tools, the authors rigorously reviewed, modified, and validated all generated text, research insights, and visual content. The authors assume full and sole responsibility for the integrity, accuracy, and originality of the final manuscript and affirm that no AI tool fulfills the role of, nor is listed as, an author.

\appendix

\section{Common gap rescaling and Gibbs preservation}
\label{app:quan}

Common gap rescaling follows from requiring an adiabatic stroke to map a Gibbs state at inverse temperature $\beta_h$ into a Gibbs state at inverse temperature $\beta_l$ without additional irreversible thermalization. Since a quantum adiabatic stroke preserves populations,
\begin{equation}
\frac{P_n}{P_m}
=
\exp[-\beta_h(E_n^h-E_m^h)]
=
\exp[-\beta_l(E_n^l-E_m^l)] .
\end{equation}
For all pairs $n,m$, this requires
\begin{equation}
E_n^l-E_m^l
=
\frac{\beta_h}{\beta_l}
(E_n^h-E_m^h),
\end{equation}
which is the homothetic condition with $\alpha=\beta_h/\beta_l$ for the reversible Carnot connection. The Otto cycle does not require this value of $\alpha$; in the Otto setting, $\alpha$ is an operating parameter constrained by the positive-work condition. The same algebra gives a level-independent multilevel homothetic Otto efficiency.

\section{TPM cumulants and moment reduction}
\label{app:moment_details}

The cumulant-generating function associated with Eq.~\eqref{eq:general_PW} is
\begin{equation}
K_W(t)
=
\ln\sum_{n=0}^{N-1}p_n^h e^{t(1-\alpha)E_n^h}
+
\ln\sum_{m=0}^{N-1}p_m^l e^{-t(1-\alpha)E_m^h},
\label{eq:general_cgf_app}
\end{equation}
with $\kappa_r^{(W)}=\partial_t^rK_W(t)|_{t=0}$. The joint work--heat distribution is
\begin{align}
P_N(W,Q_h)
&=
\sum_{n,m=0}^{N-1}p_n^h p_m^l
\delta\!\left[W-(1-\alpha)(E_n^h-E_m^h)\right]
\nonumber\\
&\quad\times
\delta\!\left[Q_h-(E_n^h-E_m^h)\right].
\label{eq:joint_WQ_app}
\end{align}
It is supported on $W=(1-\alpha)Q_h$. This support gives the scale-invariant quasistatic proportionality between work and input-heat cumulants used in fluctuation-bound analyses of quantum Otto cycles~\cite{SaryalAgarwalla2021}; the finite-ladder form is given in Appendix~\ref{app:ladder_details}.

The first two moments follow by expanding Eq.~\eqref{eq:general_PW}:
\begin{align}
\langle W\rangle_N
&=
(1-\alpha)
\left[
\sum_n p_n^hE_n^h-\sum_m p_m^lE_m^h
\right]
\nonumber\\
&=
(1-\alpha)(\mu_h-\mu_l^{(h)}),
\end{align}
and
\begin{equation}
\langle W^2\rangle_N
=
(1-\alpha)^2
\left[
M_{2,h}+M_{2,l}^{(h)}-2\mu_h\mu_l^{(h)}
\right].
\end{equation}
Subtracting $\langle W\rangle_N^2$ gives Eq.~\eqref{eq:var_general}.

\section{Finite-ladder cumulants and limiting expansions}
\label{app:ladder_details}

For the work index $k=n-m$, the cumulant-generating function is
\begin{equation}
K_N(t)
=
\ln\langle e^{tk}\rangle
=
\ln\frac{Z_N(z_h-t)}{Z_N(z_h)}
+
\ln\frac{Z_N(z_l+t)}{Z_N(z_l)}.
\label{eq:cumulant_generator}
\end{equation}
Thus $\kappa_r^{(k)}=\partial_t^rK_N(t)|_{t=0}$ and
\begin{equation}
\kappa_r^{(W)}
=
\left[(1-\alpha)\epsilon\right]^r
\kappa_r^{(k)}.
\label{eq:work_cumulants_appendix}
\end{equation}
Since the quasistatic homothetic cycle also has $Q_h=\epsilon k$, the hot-heat cumulants obey
\begin{equation}
\kappa_r^{(Q_h)}
=
\epsilon^r\kappa_r^{(k)}.
\label{eq:heat_cumulants_appendix}
\end{equation}
Consequently,
\begin{equation}
\kappa_r^{(W)}
=
(1-\alpha)^r
\kappa_r^{(Q_h)}.
\label{eq:work_heat_cumulant_relation}
\end{equation}
This is the finite-ladder form of the scale-invariant quasistatic cumulant proportionality. With the opposite sign convention, where positive work denotes work performed on the working fluid, the odd work cumulants acquire the corresponding sign change. Here $W>0$ denotes extracted work.

In particular,
\begin{equation}
\kappa_1^{(k)}
=
\nu_N(z_h)-\nu_N(z_l),
\qquad
\kappa_2^{(k)}
=
v_N(z_h)+v_N(z_l).
\end{equation}
The third cumulant is
\begin{equation}
\kappa_3^{(k)}
=
-\partial_z^3\ln Z_N(z_h)
+
\partial_z^3\ln Z_N(z_l).
\label{eq:kappa3_uniform}
\end{equation}
For $r=z_l/z_h>1$, taking $N\to\infty$ before the high-temperature limit gives Eq.~\eqref{eq:gamma1_HT_osc}.

At low temperature, $\nu_N(z)\simeq e^{-z}$ and $v_N(z)\simeq e^{-z}$ up to $O(e^{-2z})$ and cutoff terms $O(e^{-Nz})$, giving Eq.~\eqref{eq:lowT_snr}. At fixed finite $N$ and $Nz\ll1$,
\begin{equation}
\nu_N(z)
\simeq
\frac{N-1}{2}-\frac{N^2-1}{12}z,
\qquad
v_N(z)
\simeq
\frac{N^2-1}{12},
\label{eq:near_uniform_nu_v}
\end{equation}
which gives Eq.~\eqref{eq:near_uniform}.

\section{Useful finite dimension and design maps}
\label{app:dimension_design}

The finite uniform ladder approaches the oscillator result once the upper cutoff lies above the thermally populated part of the spectrum. Here we convert that observation into a practical diagnostic: for fixed reservoir parameters, the useful dimension is the smallest ladder size beyond which additional levels change a chosen diagnostic by less than a prescribed tolerance.

For a uniform homothetic ladder, define the dimensionless mean-work factor
\begin{equation}
\bar W_N(z_h,z_l)
=
\nu_N(z_h)-\nu_N(z_l),
\label{eq:app_Wbar_N}
\end{equation}
so that
\begin{equation}
\langle W\rangle_N=(1-\alpha)\epsilon \bar W_N .
\end{equation}
The corresponding work reliability is
\begin{equation}
\mathcal R_N(z_h,z_l)
=
\frac{\nu_N(z_h)-\nu_N(z_l)}
{\sqrt{v_N(z_h)+v_N(z_l)}} .
\label{eq:app_R_N_design}
\end{equation}
The oscillator-limit quantities are obtained from
\begin{align}
\nu_N(z)&\to \nu_\infty(z)=\frac{1}{\e^z-1},
\nonumber\\
v_N(z)&\to v_\infty(z)=\frac{\e^z}{(\e^z-1)^2}.
\label{eq:app_osc_limit_design}
\end{align}
We denote these limits by $\bar W_\infty$ and $\mathcal R_\infty$.

\begin{figure*}[t]
\centering
\includegraphics[width=0.8\linewidth]{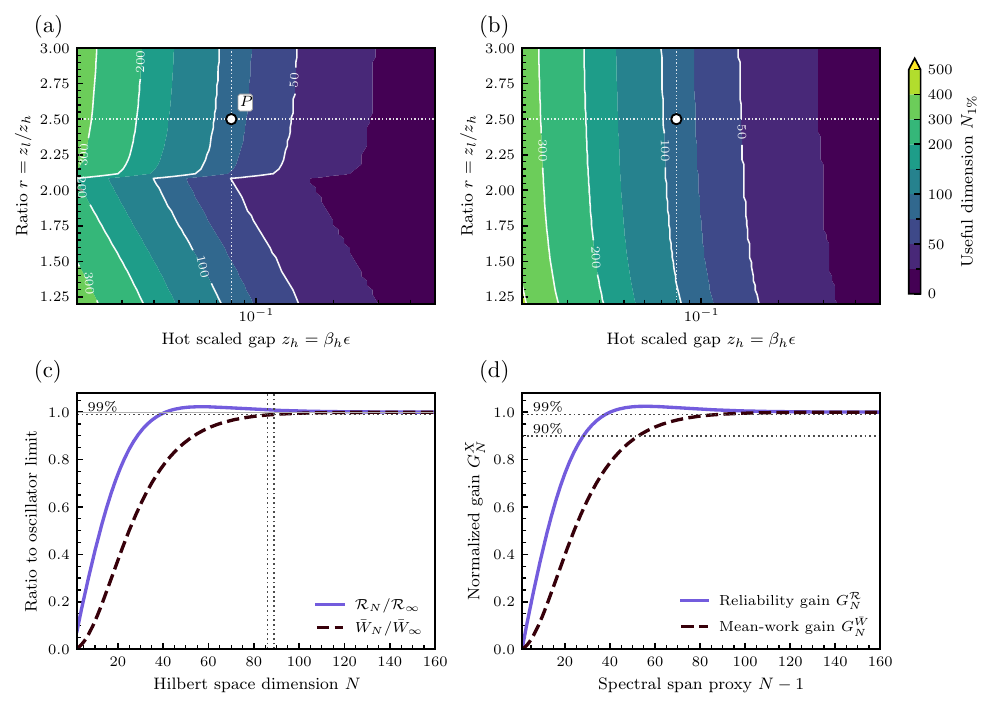}
\caption{
Finite dimension as an engineering design resource.
(a) Smallest ladder dimension $N_{1\%}^{\mathcal R}$ for which the finite-$N$ work reliability remains within $1\%$ of the oscillator value for all larger scanned cutoffs up to $N_{\rm scan}=500$, as a function of the hot scaled gap $z_h$ and the ratio $r=z_l/z_h$.
(b) Corresponding dimension $N_{1\%}^{\bar W}$ for the dimensionless mean-work factor $\bar W_N=\nu_N(z_h)-\nu_N(z_l)$, using the same finite-scan tail criterion. The marked point is $z_h=0.08$, $z_l=0.20$.
(c) Representative convergence of $\mathcal R_N/\mathcal R_\infty$ and $\bar W_N/\bar W_\infty$ at the marked point.
(d) Fraction of the available qubit-to-oscillator gain recovered as a function of the spectral-span proxy $N-1$.
Adding levels changes the reliability and mean-work diagnostics appreciably only until the cutoff exceeds the thermally active range.
}
\label{fig:dimension_design}
\end{figure*}

A finite ladder can cross the oscillator value at an isolated cutoff. Such crossings do not provide a stable cutoff criterion. We therefore use a finite-scan tail distance
\begin{equation}
\Delta_{N,N_{\rm scan}}^X
=
\max_{N\le M\le N_{\rm scan}}
\frac{|X_M-X_\infty|}{|X_\infty|},
\qquad
X\in\{\mathcal R,\bar W\}.
\label{eq:app_tail_distance_design}
\end{equation}
The useful dimension at tolerance $\delta$ is
\begin{equation}
N_\delta^X
=
\min\{N:\Delta_{N,N_{\rm scan}}^X<\delta\},
\qquad
X\in\{\mathcal R,\bar W\}.
\label{eq:app_useful_dimension}
\end{equation}
All design maps in Fig.~\ref{fig:dimension_design} use $\delta=0.01$ and a maximum scanned dimension $N_{\rm scan}=500$. The criterion therefore requires the finite ladder to remain within one percent of the oscillator value for every larger cutoff up to $N_{\rm scan}$, rather than merely touching the oscillator value at a single dimension.

The fraction of the available qubit-to-oscillator improvement recovered at finite $N$ is
\begin{equation}
G_N^X = \frac{X_N-X_2}{X_\infty-X_2}.
\label{eq:app_gain_fraction}
\end{equation}
The gain fraction is diagnostic dependent and need not be monotonic for arbitrary parameters. In the regimes plotted below, it quantifies diminishing returns: once $G_N^X$ is close to one, further increasing the Hilbert-space dimension changes the selected diagnostic only weakly.

The numerical evaluation uses the stable finite-$N$ forms
\begin{equation}
\nu_N(z) = \frac{q}{1-q} - \frac{Nq^N}{1-q^N}, \qquad q=\e^{-z},
\label{eq:app_nu_stable}
\end{equation}
and
\begin{equation}
v_N(z)
=
\frac{q}{(1-q)^2}
-
\frac{N^2q^N}{(1-q^N)^2}.
\label{eq:app_v_stable}
\end{equation}
These expressions are algebraically equivalent to Eqs.~\eqref{eq:nuN} and~\eqref{eq:vN}, but avoid numerical overflow when $Nz$ is large. This stabilization is needed in scans that combine large cutoffs with moderately large $z_h$.

Figure~\ref{fig:dimension_design} gives the useful-dimension diagnostics. Panels (a) and (b) show $N_{1\%}^{\mathcal R}$ and $N_{1\%}^{\bar W}$ as functions of the hot scaled gap $z_h$ and the ratio
\begin{equation}
r=\frac{z_l}{z_h}>1 .
\end{equation}
Both maps show that the required dimension grows as $z_h$ decreases, because the thermally occupied tail broadens. The ratio $r$ changes the separation between the two endpoint distributions and shifts the quantitative tolerance threshold, while the dominant scaling is set by whether the cutoff lies above the thermally active range.

Panels (c) and (d) use the representative point $z_h=0.08$ and $z_l=0.20$. Panel (c) compares $\mathcal R_N/\mathcal R_\infty$ and $\bar W_N/\bar W_\infty$ as functions of $N$. Panel (d) shows the corresponding gain fractions $G_N^{\mathcal R}$ and $G_N^{\bar W}$ as functions of the spectral-span proxy $N-1$. This proxy is not an energetic cost model; it records the range of level indices that must be controlled in the finite ladder. The gain curves saturate once the cutoff exceeds the occupied thermal tail.

The useful-dimension criterion is a cutoff diagnostic, not a universal cost-benefit theorem. The experimental cost of increasing $N$ is platform dependent: in a qudit it may involve leakage control and spectral addressability, whereas in an oscillator it corresponds to controlling truncation errors over a broader thermal tail. The finite sums support the narrower conclusion that finite Hilbert-space support matters until the thermally populated part of the oscillator distribution is resolved; beyond that point the finite ladder is oscillator-like for the chosen diagnostic and tolerance.

\section{Stationary-cycle derivation for incomplete isochores}
\label{app:incomplete_details}

We adopt a cold-to-cold convention. Let $i$ denote the level immediately after the cold isochore, $j$ the level after the hot isochore, and $k$ the level after the next cold isochore. Since the quasistatic strokes preserve the level label,
\begin{equation}
W=(1-\alpha)\epsilon(j-i),
\qquad
Q_h=\epsilon(j-i),
\qquad
Q_l=\alpha\epsilon(k-j).
\end{equation}
The stationary-cycle trajectory probability is
\begin{equation}
\mathbb P(i,j,k)=q_i^lR_h(j\mid i)R_l(k\mid j).
\label{eq:trajectory_probability_partial_app}
\end{equation}
Summing over $k$ gives Eq.~\eqref{eq:P_lambda}. The mean work index is
\begin{equation}
\langle \Delta\rangle_\lambda
=
\sum_{i,j}q_i^lR_h(j\mid i)(j-i)
=
\frac{\lambda_h\lambda_l}{D_\lambda}[\nu_N(z_h)-\nu_N(z_l)],
\end{equation}
where $\Delta=j-i$. The second moment is
\begin{equation}
\langle \Delta^2\rangle_\lambda
=
\lambda_h\left[v_h+V_l+(\nu_h-\bar n_l)^2\right],
\end{equation}
with
\begin{align} \nonumber
\nu_h&=\nu_N(z_h),
\qquad
v_h=v_N(z_h),
\\
\bar n_l&=\sum_i i q_i^l,
\qquad
V_l=\sum_iq_i^l(i-\bar n_l)^2 .    
\end{align}
The corresponding reliability is
\begin{equation}
\mathcal R_{N,\lambda}
=
\frac{\langle\Delta\rangle_\lambda}
{\sqrt{\langle\Delta^2\rangle_\lambda-\langle\Delta\rangle_\lambda^2}}.
\end{equation}
For diagonal states,
\begin{equation}
D_{\rm KL}(q^s\Vert p^s)=\beta_s[F_s(q^s)-F_s(p^s)],
\qquad
s=h,l,
\end{equation}
so $\mathcal A_{\rm diag}$ is the sum of the endpoint nonequilibrium free-energy excesses in thermal units.

\section{Finite-time transition-matrix details}
\label{app:finite_time_details}

Finite-time trajectory sums used in Sec.~\ref{sec:finite_time} are collected here. The reduced quantities entering the main text are the drift term $D_{\rm jump}$, the direct jump-spread term $V_{\rm jump}$, the covariance correction $V_{\rm cov}$, and the harmonic sudden-switch friction term controlled by $Q^*$. The final subsection gives an auxiliary noncommuting-endpoint check of the transition-matrix calculation; it is outside the exact homothetic finite-time benchmark.

\subsection{Trajectory sums and conditioned efficiency}

A finite-time trajectory is
\[
\gamma=(n,k,m,j),
\]
with probability
\begin{equation}
\Pi_\gamma
=
p_n^h T^{\rm e}_{k|n} p_m^l T^{\rm c}_{j|m}.
\label{eq:app_ft_weight}
\end{equation}
The stroke works and heats are
\begin{align}
W_{\rm e}(n,k)&=E_n^h-E_k^l,
&
W_{\rm c}(m,j)&=E_m^l-E_j^h, \\
Q_h(n,j)&=E_n^h-E_j^h,
&
Q_l(m,k)&=E_m^l-E_k^l .
\end{align}
The total extracted work is
\begin{equation}
W_\gamma
=
E_n^h-E_k^l+E_m^l-E_j^h .
\label{eq:app_ft_total_work}
\end{equation}
For trajectories with $Q_h(\gamma)\ne0$,
\begin{equation}
\eta_\gamma
=
\frac{W_\gamma}{Q_h(\gamma)}
=
1+\frac{Q_l(\gamma)}{Q_h(\gamma)} .
\label{eq:app_ft_eta}
\end{equation}
For homothetic spectra $E_n^l=\alpha E_n^h$,
\begin{equation}
\eta_\gamma
=
1-\alpha
\frac{E_k^h-E_m^h}{E_n^h-E_j^h}.
\label{eq:app_ft_eta_homothetic}
\end{equation}
For a uniform ladder $E_n^h=n\epsilon_h$, this becomes
\begin{equation}
\eta_\gamma
=
1-\alpha\frac{k-m}{n-j}.
\label{eq:app_ft_eta_uniform}
\end{equation}
The adiabatic case $k=n$, $j=m$ therefore gives $\eta_\gamma=1-\alpha$ for every trajectory with nonzero absorbed hot heat.

For any conditioned trajectory set $\Gamma$,
\begin{align}
\mathcal N_\Gamma
&=
\sum_{\gamma\in\Gamma}\Pi_\gamma,
\label{eq:app_ft_conditioned_norm}
\\
\mathcal P_\Gamma(\eta)
&=
\frac{1}{\mathcal N_\Gamma}
\sum_{\gamma\in\Gamma}
\Pi_\gamma
\delta\left(\eta-\frac{W_\gamma}{Q_h(\gamma)}\right).
\label{eq:app_ft_conditioned_eta_dist}
\end{align}
The heat-engine-conditioned set used in the diagnostics is
\[
\Gamma_{\rm HE}
=
\{\gamma:Q_h(\gamma)>0,\ Q_l(\gamma)<0,\ W_\gamma>0\}.
\]
Its moments are
\begin{equation}
\langle\eta^r\rangle_\Gamma
=
\frac{1}{\mathcal N_\Gamma}
\sum_{\gamma\in\Gamma}
\Pi_\gamma
\left[
\frac{W_\gamma}{Q_h(\gamma)}
\right]^r .
\label{eq:app_ft_conditioned_eta_moments}
\end{equation}

\subsection{Jump decomposition and work variance}

For a uniform homothetic ladder,
\[
E_n^h=n\epsilon_h,
\qquad
E_n^l=\alpha n\epsilon_h,
\]
define
\[
d_{\rm e}=k-n,
\qquad
d_{\rm c}=j-m .
\]
The trajectory work can then be written as
\begin{equation}
\frac{W_\gamma}{\epsilon_h}
=
(1-\alpha)(n-m)-\alpha d_{\rm e}-d_{\rm c}.
\label{eq:app_ft_jump_work}
\end{equation}
Averaging over the trajectory ensemble gives
\begin{align}
\langle W\rangle
&=
\langle W\rangle_{\rm ad}
-
\epsilon_hD_{\rm jump},
\nonumber\\
D_{\rm jump}
&=
\alpha\langle d_{\rm e}\rangle_h
+
\langle d_{\rm c}\rangle_l ,
\label{eq:app_ft_Djump}
\end{align}
where
\begin{align}
\langle d_{\rm e}\rangle_h
&=
\sum_{n,k}
p_n^h T^{\rm e}_{k|n}(k-n), \\
\langle d_{\rm c}\rangle_l
&=
\sum_{m,j}
p_m^l T^{\rm c}_{j|m}(j-m).
\end{align}

For the variance, write
\[
\frac{W_\gamma}{\epsilon_h}
=
\underbrace{(1-\alpha)n-\alpha d_{\rm e}}_{X_{\rm e}}
+
\underbrace{-(1-\alpha)m-d_{\rm c}}_{X_{\rm c}} .
\]
The variables $X_{\rm e}$ and $X_{\rm c}$ are independent because the isochores fully reset the populations. Hence
\begin{align}
\frac{\sigma_W^2}{\epsilon_h^2}
&=
(1-\alpha)^2(v_h+v_l)
+
\alpha^2{\rm Var}_h(d_{\rm e})
+
{\rm Var}_l(d_{\rm c})
\nonumber \\
&\quad
-
2\alpha(1-\alpha)
{\rm Cov}_h(n,d_{\rm e})
+
2(1-\alpha)
{\rm Cov}_l(m,d_{\rm c}) .
\label{eq:app_ft_jump_variance_full}
\end{align}
The direct jump-induced contribution is
\begin{equation}
V_{\rm jump}
=
\alpha^2{\rm Var}_h(d_{\rm e})
+
{\rm Var}_l(d_{\rm c}),
\label{eq:app_ft_Vjump}
\end{equation}
and the covariance correction is
\begin{equation}
V_{\rm cov}
=
-2\alpha(1-\alpha){\rm Cov}_h(n,d_{\rm e})
+
2(1-\alpha){\rm Cov}_l(m,d_{\rm c}) .
\label{eq:app_ft_Vcov}
\end{equation}
Combining Eqs.~\eqref{eq:app_ft_jump_variance_full}--\eqref{eq:app_ft_Vcov} gives the reliability ratio used in the main text,
\begin{equation}
\frac{\mathcal{R}_W}{\mathcal{R}_{W,\rm ad}}
=
\frac{
1-D_{\rm jump}/[(1-\alpha)(\nu_h-\nu_l)]
}{
\sqrt{
1+(V_{\rm jump}+V_{\rm cov})/[(1-\alpha)^2(v_h+v_l)]
}
}.
\label{eq:app_ft_snr_ratio}
\end{equation}
This expression is exact under complete thermalization for the uniform homothetic ladder once the transition matrices are specified.

Near the adiabatic limit, the transition matrices may be expanded as
\begin{align}
T^{\rm e}_{k|n}
&=
\delta_{kn}
+
\lambda_{\rm e}R^{\rm e}_{k|n}
+
O(\lambda_{\rm e}^2),
\nonumber\\
T^{\rm c}_{j|m}
&=
\delta_{jm}
+
\lambda_{\rm c}R^{\rm c}_{j|m}
+
O(\lambda_{\rm c}^2),
\label{eq:app_ft_weak_T}
\end{align}
with column sums of $R^{\rm e}$ and $R^{\rm c}$ equal to zero. The first-order mean-work correction is
\begin{equation}
\delta\langle W\rangle^{(1)}
=
-\alpha\lambda_{\rm e}
\langle\Delta_{\rm e}\rangle_h
-
\lambda_{\rm c}
\langle\Delta_{\rm c}\rangle_l ,
\label{eq:app_ft_weak_mean_correction}
\end{equation}
where
\begin{align}
\Delta_{\rm e}(n)
&=
\sum_k R^{\rm e}_{k|n}(E_k^h-E_n^h),
\nonumber\\
\Delta_{\rm c}(m)
&=
\sum_j R^{\rm c}_{j|m}(E_j^h-E_m^h).
\end{align}
For a uniform ladder this weak-transition expansion is equivalently encoded by $D_{\rm jump}=O(\lambda)$ and $V_{\rm jump}=O(\lambda)$.

\subsection{Qubit boundary}

For $N=2$, a symmetric transition matrix is determined by a single flip probability $p$,
\begin{equation}
T
=
\begin{pmatrix}
1-p & p\\
p & 1-p
\end{pmatrix}.
\label{eq:app_ft_qubit_T}
\end{equation}
For identical expansion and compression matrices, the weighted nonadiabaticity is $\mathcal A=p$. The drift terms are
\[
\langle d_{\rm e}\rangle_h=p(p_0^h-p_1^h),
\qquad
\langle d_{\rm c}\rangle_l=p(p_0^l-p_1^l),
\]
so
\begin{equation}
\frac{\langle W\rangle_{N=2}}{\epsilon_h}
=
(1-\alpha)(p_1^h-p_1^l)
-
p\left[
\alpha(p_0^h-p_1^h)
+
(p_0^l-p_1^l)
\right].
\label{eq:app_ft_qubit_mean}
\end{equation}
The qubit is the single-flip-channel limit of the general jump-drift formula.

\subsection{Harmonic sudden switch}

For the harmonic oscillator,
\[
H_\omega
=
\frac{p^2}{2}
+
\frac{\omega^2x^2}{2}.
\]
A sudden switch $\omega_h\to\omega_l=\alpha\omega_h$ produces the transition matrix
\begin{equation}
T^{\rm ss}_{k|n}
=
|\langle k;\omega_l|n;\omega_h\rangle|^2.
\label{eq:app_ft_harmonic_T}
\end{equation}
The standard sudden-switch nonadiabaticity factor is
\begin{equation}
Q^*
=
\frac{\omega_h^2+\omega_l^2}{2\omega_h\omega_l}
=
\frac{1+\alpha^2}{2\alpha}.
\label{eq:app_ft_Qstar}
\end{equation}
The conditional mean final level is  

\begin{equation}
\overline{k|n}
=
Q^*\left(n+\frac12\right)-\frac12 .
\label{eq:app_ft_Qstar_conditional_mean}
\end{equation}
Therefore
\begin{equation}
\langle W\rangle_{\rm ss}
=
\langle W\rangle_{\rm ad}
-
W_{\rm fric},
\label{eq:app_ft_sudden_mean}
\end{equation}
where
\begin{equation}
\langle W\rangle_{\rm ad}
=
(1-\alpha)\epsilon_h(\nu_h-\nu_l),
\label{eq:app_ft_sudden_ad_work}
\end{equation}
and
\begin{equation}
W_{\rm fric}
=
\epsilon_h(Q^*-1)
\left[
\alpha\left(\nu_h+\frac12\right)
+
\left(\nu_l+\frac12\right)
\right].
\label{eq:app_ft_sudden_friction}
\end{equation}

The squeezed-transition variance is
\begin{equation}
{\rm Var}(k|n)
=
\frac12(n^2+n+1)\left[(Q^*)^2-1\right].
\label{eq:app_ft_conditional_variance}
\end{equation}
Hence
\begin{align}
\frac{\sigma_{W,\rm ss}^2}{\epsilon_h^2}
&=
(1-\alpha Q^*)^2v_h
+
(\alpha-Q^*)^2v_l
\nonumber \\
&\quad
+
\frac12\left[(Q^*)^2-1\right]
\left[
\alpha^2\langle n^2+n+1\rangle_h
+
\langle m^2+m+1\rangle_l
\right].
\label{eq:app_ft_sudden_variance}
\end{align}
These equations underlie Fig.~\ref{fig:harmonic_Qstar_sudden_switch}. The $Q^*$ formulas describe the harmonic oscillator boundary; the finite matrices used in Fig.~\ref{fig:harmonic_Qstar_sudden_switch} are numerical oscillator cutoffs, not physical finite uniform ladders.

\subsection{Auxiliary noncommuting-endpoint check}

A sudden gap change of the strict uniform homothetic ladder is commuting: $H_h=\epsilon_h n_N$ and $H_l=\alpha\epsilon_h n_N$ share the same eigenvectors. Therefore a sudden gap rescaling alone gives $T_{k|n}=\delta_{kn}$. Nontrivial sudden-switch transitions in a finite ladder require noncommuting endpoint Hamiltonians. One such auxiliary check uses
\begin{equation}
H_{\rm pre}^{(N)}
=
\epsilon_h n_N+gV_N,
\qquad
H_{\rm post}^{(N)}
=
\alpha\epsilon_h n_N+gV_N,
\label{eq:app_noncommuting_sudden_endpoints}
\end{equation}
where
\begin{equation}
V_N
=
\frac{a_N+a_N^\dagger}{\|a_N+a_N^\dagger\|_2}.
\label{eq:app_noncommuting_sudden_VN}
\end{equation}
For $N>2$, these endpoint spectra are generally not related by common gap rescaling. The calculation is therefore an auxiliary transition-matrix check, not part of the exact homothetic finite-time benchmark and not a substitute for the quasistatic weak-nonhomothety expansion in Sec.~\ref{sec:nonhom}.

Let
\begin{align}
H_{\rm pre}^{(N)}|\psi_n^{\rm pre}\rangle
&=
E_n^{\rm pre}|\psi_n^{\rm pre}\rangle, \\
H_{\rm post}^{(N)}|\psi_k^{\rm post}\rangle
&=
E_k^{\rm post}|\psi_k^{\rm post}\rangle ,
\end{align}
with the ground energies shifted to zero. The sudden-switch transition matrix is
\begin{equation}
T^{(N)}_{k|n}
=
\left|
\langle\psi_k^{\rm post}|\psi_n^{\rm pre}\rangle
\right|^2 .
\label{eq:app_noncommuting_sudden_T}
\end{equation}
The TPM work values are
\begin{equation}
W_{\rm e}(n,k)=E_n^{\rm pre}-E_k^{\rm post},
\qquad
W_{\rm c}(m,j)=E_m^{\rm post}-E_j^{\rm pre}.
\label{eq:app_noncommuting_sudden_work_values}
\end{equation}
The work moments follow from the same trajectory sums as Eq.~\eqref{eq:app_ft_weight}, with $E^h\to E^{\rm pre}$, $E^l\to E^{\rm post}$, $T^{\rm e}=T^{(N)}$, and $T^{\rm c}=(T^{(N)})^{\mathsf T}$.

For $N=2$,
\[
n_2=
\begin{pmatrix}
0&0\\
0&1
\end{pmatrix},
\qquad
V_2=\sigma_x,
\]
so
\begin{equation}
H(\epsilon,g)
=
\begin{pmatrix}
0&g\\
g&\epsilon
\end{pmatrix}
=
\frac{\epsilon}{2}\mathbb I
+
g\sigma_x
-
\frac{\epsilon}{2}\sigma_z .
\label{eq:app_noncommuting_qubit_H}
\end{equation}
The shifted gap is
\begin{equation}
\Omega(\epsilon,g)=\sqrt{\epsilon^2+4g^2}.
\label{eq:app_noncommuting_qubit_gap}
\end{equation}
Writing $\epsilon_l=\alpha\epsilon_h$,
\[
\Omega_h=\sqrt{\epsilon_h^2+4g^2},
\qquad
\Omega_l=\sqrt{\epsilon_l^2+4g^2}.
\]
The two endpoint Bloch fields are
\[
\mathbf b_h=\left(g,0,-\frac{\epsilon_h}{2}\right),
\qquad
\mathbf b_l=\left(g,0,-\frac{\epsilon_l}{2}\right),
\]
and the sudden-switch flip probability is
\begin{equation}
p_{\rm sw}
=
\frac12
\left[
1-
\frac{4g^2+\epsilon_h\epsilon_l}
{\sqrt{\epsilon_h^2+4g^2}\sqrt{\epsilon_l^2+4g^2}}
\right].
\label{eq:app_noncommuting_qubit_flip}
\end{equation}
Thus
\begin{equation}
T^{(2)}
=
\begin{pmatrix}
1-p_{\rm sw} & p_{\rm sw}\\
p_{\rm sw} & 1-p_{\rm sw}
\end{pmatrix}.
\label{eq:app_noncommuting_qubit_T}
\end{equation}
This reduces to the identity when $g=0$ or $\alpha=1$.

For $N>2$, the departure from common gap rescaling is quantified by the spread of adjacent gap ratios,
\begin{equation}
\Delta_{\rm hom}^{(N)}
=
{\rm ptp}_{q}
\left[
\frac{E_{q+1}^{\rm post}-E_q^{\rm post}}
{E_{q+1}^{\rm pre}-E_q^{\rm pre}}
\right],
\label{eq:app_noncommuting_gap_ratio_spread}
\end{equation}
where ${\rm ptp}$ denotes maximum minus minimum over adjacent gaps. 

For finite $N$, using the same nominal $\alpha$ does not by itself guarantee homothetic behavior. Exact homothety requires all populated gap ratios to be equal. If the ratios vary across the spectrum, different trajectories acquire different effective compression ratios, so the work reliability and stochastic-efficiency distribution can change even when the same nominal $\alpha$ is used. The direction of the change is spectrum dependent; it is controlled by the thermally occupied distorted gaps and by their covariance with the endpoint Gibbs weights.

\begin{figure*}[t]
\centering
\includegraphics[width=0.7\linewidth]{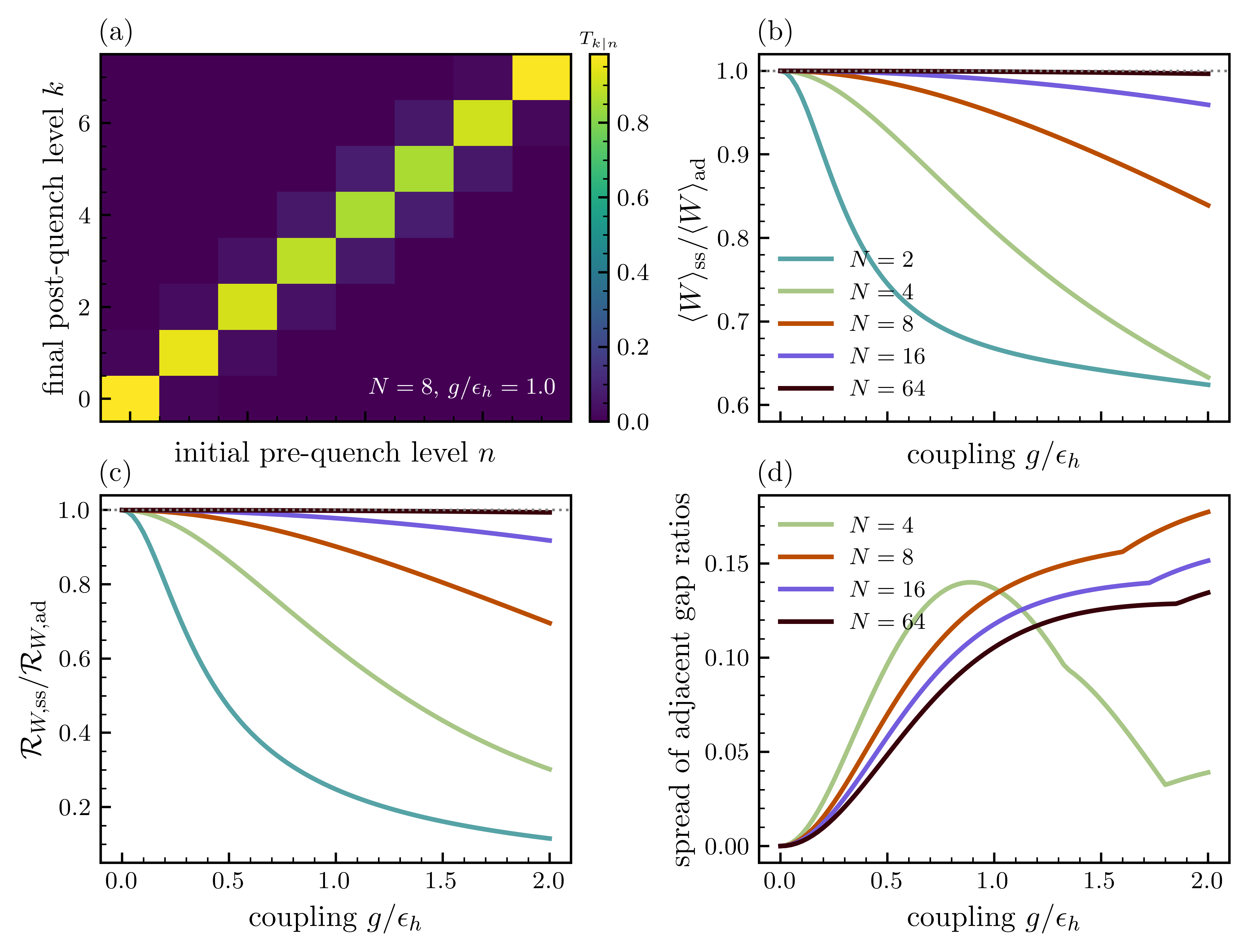}
\caption{
Finite-$N$ noncommuting sudden-switch extension. The endpoint Hamiltonians are $H_{\rm pre}^{(N)}=\epsilon_h n_N+gV_N$ and $H_{\rm post}^{(N)}=\alpha\epsilon_h n_N+gV_N$, with $V_N=(a_N+a_N^\dagger)/\|a_N+a_N^\dagger\|_2$. The parameters are $\alpha=0.65$, $z_h=0.08$, and $z_l=0.20$. (a) Sudden-switch transition matrix $T^{(N)}_{k|n}=|\langle\psi_k^{\rm post}|\psi_n^{\rm pre}\rangle|^2$ for $N=8$ and $g/\epsilon_h=1$. (b) Mean-work ratio $\langle W\rangle_{\rm ss}/\langle W\rangle_{\rm ad}$ versus $g/\epsilon_h$. (c) Work-reliability ratio $\mathcal R_{W,\rm ss}/\mathcal R_{W,\rm ad}$. Small Hilbert spaces are more sensitive to the sudden endpoint mismatch than large cutoffs. (d) Endpoint nonhomothety diagnostic $\Delta_{\rm hom}^{(N)}$, defined in Eq.~\eqref{eq:app_noncommuting_gap_ratio_spread}. For $N=2$, any shifted two-level spectrum is homothetic in the trivial two-level sense, so this diagnostic is not shown.
}
\label{fig:app_noncommuting_sudden_switch}
\end{figure*}

In the qubit-like avoided-crossing limit, the sudden endpoint mismatch produces a stronger reduction of reliability than in larger cutoffs. The large-$N$ curves remain close to the adiabatic reference over the plotted range because the chosen normalization of $V_N$ weakens the effective spectral distortion per accessible level. Panel (d) gives the corresponding endpoint nonhomothety diagnostic.

\section{One-cycle, repeated-cycle, and limit-cycle interpretation}
\label{app:cycle_interpretation}

The trajectory weight in Eq.~\eqref{eq:app_ft_weight} assumes complete thermalization on both isochores. The working medium is prepared in $p^h$ before each expansion stroke and in $p^l$ before each compression stroke. Consecutive cycles are therefore statistically independent. The one-cycle formulas are also the per-cycle statistics of the stationary periodic operation under complete thermalization.

For $M$ repeated cycles,
\begin{equation}
W_M=\sum_{r=1}^M W^{(r)},
\label{eq:app_M_cycles_sum}
\end{equation}
where the random variables $W^{(r)}$ are independent and identically distributed. Hence
\begin{align}
\langle W_M\rangle &= M\langle W\rangle, \label{eq:app_M_cycles_mean} \\
\sigma^2_{W_M} &= M\sigma_W^2, \label{eq:app_M_cycles_var} \\
{\mathcal{R}}_{W_M} &= \sqrt{M}\,{\mathcal{R}}_W . \label{eq:app_M_cycles_reliability}
\end{align}
The same scaling holds for the adiabatic reference, so
\begin{equation}
\frac{{\mathcal{R}}_{W_M}}{{\mathcal{R}}_{W_M,\rm ad}}
=
\frac{{\mathcal{R}}_W}{{\mathcal{R}}_{W,\rm ad}} .
\label{eq:app_M_cycles_ratio}
\end{equation}

For a deterministic cycle duration $\tau_{\rm cyc}$, power is a rescaled work variable. For a single cycle,
\begin{equation}
P_\gamma=\frac{W_\gamma}{\tau_{\rm cyc}} .
\label{eq:app_single_cycle_power}
\end{equation}
Therefore
\begin{equation}
\langle P\rangle=\frac{\langle W\rangle}{\tau_{\rm cyc}},
\qquad
\sigma_P=\frac{\sigma_W}{\tau_{\rm cyc}},
\qquad
{\mathcal R}_P
=
\frac{\langle P\rangle}{\sigma_P}
=
{\mathcal R}_W .
\label{eq:app_power_single_cycle_scaling}
\end{equation}
For $M$ independent cycles with the same deterministic duration, the time-averaged power is
\begin{equation}
P_M=\frac{W_M}{M\tau_{\rm cyc}} .
\label{eq:app_M_cycle_power}
\end{equation}
Its mean and variance are
\begin{equation}
\langle P_M\rangle=\frac{\langle W\rangle}{\tau_{\rm cyc}},
\qquad
\sigma^2_{P_M}=\frac{\sigma_W^2}{M\tau_{\rm cyc}^2},
\label{eq:app_M_cycle_power_mean_var}
\end{equation}
and
\begin{equation}
{\mathcal R}_{P_M}
=
\sqrt{M}\,{\mathcal R}_W .
\label{eq:app_M_cycle_power_reliability}
\end{equation}
Thus deterministic-time power statistics contain no information beyond the corresponding work statistics. Stochastic cycle durations, work-duration correlations, or timing noise require a joint trajectory distribution for work and time.

With incomplete isochores, the factorization into independent cycles no longer holds. The finite-time transition matrices must be combined with the relaxation maps to form the full cycle map. Single-cycle statistics are then evaluated from the stationary periodic populations, and multi-cycle variances generally contain inter-cycle covariances.

\section{Numerical checks}
\label{app:numerics}

\begin{table}[b]
\caption{
Cutoff stabilization of the harmonic sudden-switch benchmark at $z_h=0.12$, $\tau=0.35$, and $\alpha=0.75$. 
The corresponding cold scaled gap is $z_l=\alpha z_h/\tau=0.257143$. 
The mean work is reported in units of the hot oscillator gap, $\langle W\rangle/\epsilon_h$. 
The nonadiabaticity $\mathcal A$ is the thermally weighted probability of not preserving the instantaneous number label over the two sudden-switch strokes, defined by Eq.~\eqref{eq:finite_time_weighted_A} using the squeezed-number transition matrices. 
The reliability $\mathcal R_W=\langle W\rangle/\sigma_W$ is dimensionless. 
The efficiency width $\sigma_\eta$ is computed from the trajectory ratio $\eta_{\rm st}=W/Q_h$ conditioned on heat-engine trajectories with $W>0$ and $Q_h>0$. 
The retained probability $P_{\rm ret}$ is the thermally weighted probability remaining inside the finite oscillator basis over the two sudden-switch strokes. 
The largest-$N$ rows show the cutoff scale required for stable finite-time diagnostics.
}
\label{tab:sudden_switch_stabilization}
\begin{ruledtabular}
\begin{tabular}{cccccc}
$N$ & $\mathcal A$ & $\langle W\rangle/\epsilon_h$ & $\mathcal R_W$ & $\sigma_\eta$ & $P_{\rm ret}$ \\
\hline
48  & 0.306559 & 0.646288 & 0.219676 & 0.253041 & 0.996896 \\
64  & 0.307274 & 0.676299 & 0.224088 & 0.252684 & 0.999221 \\
80  & 0.307380 & 0.682897 & 0.224806 & 0.252631 & 0.999813 \\
96  & 0.307397 & 0.684258 & 0.224912 & 0.252624 & 0.999956 \\
128 & 0.307399 & 0.684587 & 0.224931 & 0.252623 & 0.999998 \\
\end{tabular}
\end{ruledtabular}
\end{table}

The harmonic sudden-switch benchmark approximates an oscillator transition matrix in a finite basis and therefore requires a cutoff check. At finite cutoff, probability can leave the retained oscillator subspace through transitions to levels above the cutoff. The relevant diagnostic is the thermally weighted retained probability for the finite-temperature cycle, rather than the worst-case leakage from highly excited edge states.

Table~\ref{tab:sudden_switch_stabilization} reports the cutoff dependence for the parameter point used in the sudden-switch benchmark. Small cutoffs are inaccurate. Relative to the $N=128$ reference value, the mean work differs by less than $0.25\%$ at $N=80$ and less than $0.05\%$ at $N=96$. The heat-engine-conditioned efficiency width is stable to better than $0.01\%$ over the same range. The retained trajectory probability increases from $0.999813$ at $N=80$ to $0.999998$ at $N=128$.

\newpage

\bibliographystyle{apsrev4-2}
	\bibliography{references}
	
\end{document}